\documentclass[conference,dvipsnames]{IEEEtran}
\IEEEoverridecommandlockouts
\usepackage{cite}
\usepackage{amsmath,amssymb,amsfonts}
\usepackage{algorithmic}
\usepackage{graphicx}
\usepackage{eurosym}
\usepackage{textcomp}
\usepackage{xcolor}
\usepackage{xargs}
\usepackage{color,listings}
\usepackage{tabularx}

\usepackage[colorinlistoftodos,prependcaption,textsize=scriptsize]{todonotes}
\def\BibTeX{{\rm B\kern-.05em{\sc i\kern-.025em b}\kern-.08em
    T\kern-.1667em\lower.7ex\hbox{E}\kern-.125emX}}

\newcommandx{\rocky}[2][1=]{\todo[linecolor=Blue,backgroundcolor=Blue!25,bordercolor=Blue,#1]{\textbf{Rocky comments: }#2}}
\newcommandx{\mitra}[2][1=]{\todo[linecolor=OliveGreen,backgroundcolor=OliveGreen!25,bordercolor=OliveGreen,#1]{\textbf{Mitra comments: }#2}}
\newcommandx{\pragyan}[2][1=]{\todo[linecolor=Red,backgroundcolor=Yellow!25,bordercolor=Red,#1]{\textbf{Pragyan comments: }#2}}
\newcommandx{\tdb}[2][1=]{\todo[linecolor=Mulberry,backgroundcolor=Orchid!25,bordercolor=Mulberry,#1]{\textbf{Travis comments: }#2}}
\newcommandx{\sepideh}[2][1=]{\todo[linecolor=CadetBlue,backgroundcolor=CadetBlue!25,bordercolor=CadetBlue,#1]{\textbf{Sepideh comments: }#2}}
\newcommandx{\personF}[2][1=]{\todo[linecolor=Bittersweet,backgroundcolor=Bittersweet!25,bordercolor=Bittersweet,#1]{\textbf{PersonF comments: }#2}}
\newcommandx{\personG}[2][1=]{\todo[linecolor=RoyalBlue,backgroundcolor=ProcessBlue!25,bordercolor=RoyalBlue,#1]{\textbf{PersonG comments: }#2}}

\newcommandx{\citeme}[2][1=]{\todo[linecolor=red,backgroundcolor=red!25,bordercolor=red,#1]{\textbf{CITE} #2}}
\newcommandx{\fillin}[2][1=]{\todo[linecolor=red,backgroundcolor=red!25,bordercolor=red,#1]{\textbf{FILL IN} #2}}

\def\BibTeX{{\rm B\kern-.05em{\sc i\kern-.025em b}\kern-.08em
    T\kern-.1667em\lower.7ex\hbox{E}\kern-.125emX}}
\begin{document}
\title{Toward Regulatory Compliance: A few-shot Learning Approach to Extract Processing Activities}
%\title{Leveraging Few-Shot Learning to Extract Records of Processing Activities}
%Regulatory Compliance: Leveraging Few-Shot Learning for Automated Generation of Records of Processing Activities
%Toward Regulatory Compliance: A few-shot Learning Approach to Process Activity Generation

\author{\IEEEauthorblockN{Pragyan K C\textsuperscript{1}, Rambod Ghandiparsi\textsuperscript{1}, Rocky Slavin\textsuperscript{1}, Sepideh Ghanavati\textsuperscript{2}, Travis Breaux\textsuperscript{3}, Mitra Bokaei Hosseini\textsuperscript{1}}
\IEEEauthorblockA{\textsuperscript{1}University of Texas at San Antonio, San Antonio, TX, USA, \textsuperscript{2}University of Maine, Orono, ME, USA, 
\\\textsuperscript{3}Carnegie Mellon University, Pittsburgh, PA, USA\\
\textit{
[pragyan.kc, rambod.ghandiparsi, rocky.slavin, 	mitra.bokaeihosseini]@utsa.edu, sepideh.ghanavati@maine.edu, breaux@cs.cmu.edu}
}

}

\maketitle

\begin{abstract}

The widespread use of mobile applications has driven the growth of the industry, with companies relying heavily on user data for services like targeted advertising and personalized offerings. In this context, privacy regulations such as the General Data Protection Regulation (GDPR) play a crucial role. One of the GDPR requirements is the maintenance of a Record of Processing Activities (RoPA) by companies. RoPA encompasses various details, including the description of data processing activities, their purposes, types of data involved, and other relevant external entities. 
Small app-developing companies face challenges in meeting such compliance requirements due to resource limitations and tight timelines. To aid these developers and prevent fines, we propose a method to generate segments of RoPA from user-authored usage scenarios using large language models (LLMs). Our method employs few-shot learning with GPT-3.5 Turbo to summarize usage scenarios and generate RoPA segments. We evaluate different factors that can affect few-shot learning performance consistency for our summarization task, including the number of examples in few-shot learning prompts, repetition, and order permutation of examples in the prompts. Our findings highlight the significant influence of the number of examples in prompts on summarization F1 scores, while demonstrating negligible variability in F1 scores across multiple prompt repetitions. Our prompts achieve successful summarization of processing activities with an average 70\% ROUGE-L F1 score. Finally, we discuss avenues for improving results through manual evaluation of the generated summaries.

\end{abstract}


%\section*{Acknowledgment}
\section{Introduction}

The widespread adoption and use of mobile applications (apps) by users have fueled the expansion of the industry across various domains and categories. In this contemporary era, app-developing companies offer services and products that heavily rely on extensive user data. The analysis and utilization of personal data have become significant drivers of growth, particularly in areas such as targeted advertising and personalized services~\cite{wang2018guileak,slavin2016toward}. To ensure that such data processing aligns with the fundamental rights and freedoms of individuals, privacy regulations such as the General Data Protection Regulation (GDPR) require companies to maintain a record of processing activities (RoPA)~\cite{gdpr}. RoPA includes various requirements (see Section~\ref{sec:background}) regarding the details of data processing activities, which involve information such as the name of the data processing activity, its description, purposes, types of data involved, and other relevant entities~\cite{huth2019using}. 

Companies with fewer than 250 employees are exempt from these record-keeping obligations if the processing is unlikely to present a risk to the rights of the data subject, no special categories of data are processed, or if the processing occurs only occasionally, as stipulated in the GDPR - Article 30(5)~\cite{gdpr}. However, in practice, this exemption is seldom applicable as the term ``only occasional'' is ambiguous, and companies process users' data regularly. Failure to keep RoPA or provide a comprehensive index to authorities can result in fines up to \euro{10.000.000} or 2\% of the total worldwide annual turnover under the GDPR - Article 83(4)(a)~\cite{gdpr}. For example, in November 2020, Vodafone Italia was fined for \euro{12.3} million because of a vast range of GDPR violations~\cite{gdprFines}. In March 2021, Vodafone Spain was issued \euro{8.15} million by the Spanish DPA. Vodafone could have prevented these fines by regularly auditing its data processing activities, establishing data processing agreements with contractors, and maintaining clear records of all relationships with third-party data processors.

Small app-developing companies (i.e., those that are significantly smaller than 250 personnel) not only struggle with these compliance requirements but also face resource limitations, have tight development timelines to compete in the market, and have limited access to legal or privacy experts to make important privacy decisions~\cite{balebako2014improving,balebako2014privacy, Alomar2022}. Further, such companies frequently overlook creating in-depth documentation during app development processes~\cite{wagenaar2018working,hess2017towards}. Recent studies show that developers, especially in small or medium-sized companies, mainly consider privacy concepts as an afterthought and are not generally familiar enough with those concepts~\cite{Alomar2022, Bednar2019, Hadar2018, ekambaranathan2021money, spiekermann2018inside, spiekermann2018understanding, balebako2014privacy, Dalela2021AMS, tahaei2021developers, Green2016DevelopersAN, Tahaei2021Springer, prybylo2024evaluating}. In addition, studies show that many developers are only familiar with a limited number of privacy-by-design approaches~\cite{TahaeiLiVaniea+2022+114+131}. A recent survey reviewed the RoPA practices of 30 public organizations, where only 7 (23\%) of the RoPA practices contained sufficient detail for the processing activities and their purposes~\cite{RoPAReport}. Therefore, there is a significant challenge for small app-developing companies to create and maintain RoPA to comply with the GDPR's processing activities requirements and avoid potential legal repercussions. 

%Review studies that have addressed the problem. 
The current research domain predominantly focuses on formally representing RoPA using knowledge bases and semantic models to enhance automatic accountability checking and querying~\cite{martinez2021ontoropa,ryan2022support}. This trend highlights the lack of emphasis on the actual creation of RoPA. Huth et al. suggested leveraging Enterprise Architecture (EA) for RoPA creation~\cite{huth2019using}. The presence of EA signifies organizational maturity, supported by adequate resources and personnel capable of sustaining EA development and maintenance. This scenario contrasts sharply with small app-developing companies, which often operate with limited personnel and resources, leading to a lack of comprehensive documentation~\cite{wagenaar2018working,hess2017towards}.

To assist such developers in creating RoPA, we propose a framework to generate segments of RoPA for existing mobile apps. To address the lack of documentation in small app-developing companies, the framework takes advantage of usage scenarios provided by users as the primary source of processing activities. 
These scenarios can be elicited from end users through simple interactions with the app. 
This approach makes such resources accessible and attractive, particularly for smaller app development companies that lack the resources for more complex elicitation techniques. 
%During process scenario creation, an analyst guides the user through a detailed process by posing questions such as "what goal do you want to achieve?", "what do you actually do to achieve the goal?", or "how does the system help to achieve your goal?".

%\mitra[inline]{maybe add a sentence in the approach that shows the example scenario contains additional information that we don't need}
Empirical research has shown that users \textit{spontaneously} provide explanatory scenarios where they describe the app inefficiencies and desired features alongside the processing activities~\cite{anton1994goal,lubars1993developing}. 
%Besides the processing activities, such scenarios also entail noise, such as inefficiencies in the current mobile app and desired user features~\cite{anton1994goal,lubars1993developing}. 
Therefore, we propose an \textit{extractive summarization method} specifically for the purpose of filtering and extracting the processing activities from the whole scenario text, which also entails app deficiencies and desired features. %Therefore, extractive summarization plays an important role in extracting the processing activities embedded in scenarios. 
We further design experiments to investigate how well large language models (LLMs) can assist with extractive summarization to generate segments of RoPA. 

The contributions of this paper are: (1) a 50-scenario corpus with summarized processing activities; (2) an empirical evaluation of utilizing the GPT-3.5 Turbo as an instance of LLMs for extractive summarization; (3) an evaluation of different parameters affecting the LLM instance in extractive summarization task; (4) a manual evaluation of summaries extracted using the LLM instance.

%\mitra[inline]{Answer these questions: (1) Why not directly ask the users to provide the business processes? (2) Why not directly use GUI code to extract business processes?}

The remainder of this paper is organized as follows: Section~\ref{sec:background} presents background \& related work; Sections~\ref{sec:approach} \& ~\ref{sec:experimentDesign} entail the approach \& experiment designs; Sections~\ref{sec:results} \& ~\ref{sec:discussion} entail results \& discussion; Sections~\ref{sec:threats_validity} \& ~\ref{sec:conclusion} contain threats to validity \& conclusion.

\section{Background and Related Work}\label{sec:background}
%We now describe concepts used in this work and related work.

\subsection{Record of Processing Activities (RoPA)}
To comply with GDPR requirements~\cite{gdpr} in maintaining records of processing activities, companies are obligated to furnish records of: (1) the data controller's name and contact; (2) the processing purposes (e.g., the processing of contact data of suppliers for order management); (3) a description of the categories of data subjects (e.g., customers, suppliers, and employees), personal data (e.g., health data), and recipients; 
%(4) the categories of  %(even by category only) 
%to whom the personal data have been or will be communicated; 
(4) the latest deadlines for the cancellation of the different categories of data; (5) a description of the technical and organizational security measures~\cite{huth2019using,voigt2017eu}.

%Establishing and maintaining GDPR compliance requires a complete overview of the organization, an understanding of processes, applications, and data flows, a vocabulary and model to abstract these concepts, and a method to obtain consistent and reproducible results~\cite{huth2019using}. 
%A recent survey reviewed the RoPA practices of 30 public organizations, where only 7 (23\%) of the RoPAs contained sufficient detail for the processing activities and their purposes~\cite{RoPAReport}. 

%Typically, organizations manually conduct data collection and maintain isolated RoPAs ~\cite{huth2019using}. 
Huth et al. demonstrated how existing Enterprise Architecture (EA) can be enhanced with the requisite information to create and sustain a RoPA. EA encompasses a cohesive framework of principles, methodologies, and models utilized in designing and implementing an enterprise's organizational structure, business processes, information systems, and infrastructure~\cite{lankhorst2009enterprise}. Owning and maintaining EA models signifies a level of maturity within an organization, with ample resources and personnel capable of contributing to the development and maintenance of EA. However, our research targets small app-developing companies with limited personnel and resources, often lacking comprehensive documentation and EA models.

Presently, the majority of RoPA are manually created and maintained, typically presented informally in Word documents or Excel files, and often shared with the public in their original formats or as PDFs~\cite{huth2019using, martinez2021ontoropa}. To address this informality in presentation, Martinez et al. introduced a RoPA knowledge graph incorporating both legal requirements and practical insights from the privacy and data protection community~\cite{martinez2021ontoropa}. This knowledge graph offers companies a formal means of presenting their RoPA, facilitating stakeholders such as legal entities and customers in reading, evaluating, and querying the information contained within RoPA.
Ryan and Brennan developed a common semantic model to represent a machine-readable RoPA, which can be used in automated accountability systems~\cite{ryan2022support}. While these works are focused on RoPA representation, we focus on creating segments of RoPA tailored for small app-developing companies that lack comprehensive documentation throughout their app development processes. 
%In essence, their research centers on formalizing the representation of RoPAs, thereby improving their readability, assessment, and query capabilities. 

%While the current research domain predominantly focuses on ontologies and conceptual models for representing RoPA~\cite{huth2019using,martinez2021ontoropa,ryan2022support}, our paper shifts its focus towards extracting summaries of processing activities for apps. 

\subsection{Legal Compliance in RE}
Data protection officers (DPOs) and privacy engineers are pivotal experts in privacy requirements engineering, particularly for companies aiming to adhere to legal regulations~\cite{herwanto2021named}.
However, small app-developing companies may find it challenging to obtain the necessary knowledge to demonstrate compliance with regulatory requirements. 
Therefore, researchers have proposed methods and tools aimed at automating the creation of artifacts and materials required for legal compliance~\cite{Sleimi2018AutomatedEO, sleimi2020, CEJAS2021, Amaral2021, sabetzadeh21}. 
For instance, Herwanto et al. proposed a tool to help development teams automatically identify assets and privacy-related entities (such as data subject, processing, and personal data entities) from user stories~\cite{herwanto2021named} and generate data flow diagrams~\cite{herwanto2022privacystory}. Ghanavati et al. developed a Legal-URN framework for extracting legal requirements and establishing and evaluating compliance~\cite{Gha13, GRD14}. 

\subsection{Scenarios}
 
Scenarios are detailed descriptions of system behaviors, often depicted as sequences of steps, typically from a user's viewpoint~\cite{anton1998representational,sutcliffe1998scenario,sutcliffe2003scenario,weidenhaupt1998scenarios}. 
Concrete scenarios are essential to an understanding of the operational concept of a system~\cite{anton1994goal,lubars1993review}. 
They encapsulate the behavior patterns of an existing system and serve to enhance comprehension of work practices and business processes~\cite{weidenhaupt1998scenarios}. Further, these scenarios represent the aggregation of interactions, each of which is a short sequence of goal-oriented activities to achieve~\cite{potts1994inquiry}. 
In a recent study, Huang et al. proposed a framework for developers to measure the privacy risk associated with the information users provide to the app using user-authored scenarios~\cite{huang2023mobile}. %This study identifies scenarios authored by app users as a significant source of privacy requirements in mobile apps to measure privacy risk. 
We follow their survey design to collect usage scenarios describing processing activities and steps user take to accomplish their specific goals.

\subsection{Extractive Summarization}

Extractive summarization extracts text segments from an input text, capturing essential information related to the \textit{key concepts} mentioned in the input text~\cite{filatova2004event,hsu2018unified}. Unlike abstractive summarization, which rephrases the information, extractive summarization preserves the original text, ensuring factual accuracy and clarity~\cite{lin2019abstractive}. This characteristic makes it particularly suitable for factual text like our work, where preserving precise details about processing activities is crucial. 
Jadhav et al. introduced SWAP-NET, a sequence-to-sequence model for extractive summarization~\cite{jadhav2018extractive}. SWAP-NET is designed to identify salient sentences and keywords within a document, combining them to create an extractive summary. Similarly, Nallapati et al. developed a method using an RNN-based binary classifier to determine whether a sentence should be included in the summary~\cite{nallapati2017summarunner}. This method relies on the sentence's content, its significance within the document, its novelty compared to previously selected sentences, and additional positional features. Expanding on these approaches, Yadav et al. proposed a textual graph-based technique~\cite{yadav2024graph}, where document sentences form the nodes of a graph, with edges representing associations between sentences. The summary is generated based on the sentence weight and the average weight of the textual graph. Diverging from the neural network and graph-based methodologies, Mishra et al. approached summarization as a question-answering task~\cite{mishra2023llm}. They leveraged LLMs to generate pseudo-labels for dialogue, which were then used to fine-tune a chat summarization model, effectively transferring knowledge from a large LLM to a more specialized, smaller model. In contrast to these works, 
%proposed by ~\cite{jadhav2018extractive}, ~\cite{nallapati2017summarunner}, ~\cite{yadav2024graph}, ~\cite{mishra2023llm} and others, 
our method conceptualizes extractive summarization as an event-argument extraction task. By leveraging LLMs, we generate summaries of scenarios that are both more precise and less cluttered with irrelevant information. %Given action verbs within the text as pivotal events, our method locates and associates the relevant arguments with these events to construct summaries. 
%\vspace{-1em}

\subsection{Large Language Models}

In recent years, the NLP community has witnessed substantial changes due to the introduction of LLMs. 
%, such as GPT, LLaMA, and T5. %ChatGPT is a pre-trained transformer language model that has recently gained vast popularity. 
In the RE community, Fantechi et al.~\cite{fantechi2023rule} evaluated the ChatGPT's ability to identify ambiguity in requirement text and compared its performance with QuARS, a traditional rule-based NLP tool~\cite{lami2004automatic}.
%\mitra[inline]{add more LLM studies published in RE last year}. 
% 2 papers added -> Requirements Modeling Aided by ChatGPT: An Experience in Embedded Systems  -- Generating Requirements Elicitation Interview Scripts with Large Language Models
Ruan et al. incorporated ChatGPT-based zero-shot learning to extract requirement models from requirement texts and compose them using predefined rules~\cite{ruan2023requirements}.
Gorer et al. used prompt engineering to generate business domain descriptions, linear scripts, and conversation pieces focused on certain types of mistakes in requirements elicitation interviews~\cite{gorer2023generating}. 
%The results from these studies suggest a promising horizon for analyzing privacy and legal documents using large language models. However, further analysis and experiments are required to identify and address the limitations of such models.

%Their study entails four documents containing an average of 57 requirements and 1,350 words. ChatGPT is queried by asking the following prompt: ``Find the ambiguities of the following software requirements document: $\langle$\textit{List of requirements in text format}$\rangle$. QuARS and chatGPT outputs are reviewed manually, and precision and recall are calculated accordingly to measure the performance. The study findings suggest that chatGPT's performance results, especially recall, vary between chat sessions with the bot. However, precision is more stable and comparable to the result of QuARS. Further, running several sessions with the same question improves the recall. In addition, ChatGPT limits the number of input tokens for any application. Therefore, additional rule-based or manual approaches are required to break a large document into smaller chunks while preserving the context. 

\section{Approach}\label{sec:approach}

\begin{figure}
  \includegraphics[width=\columnwidth, keepaspectratio]{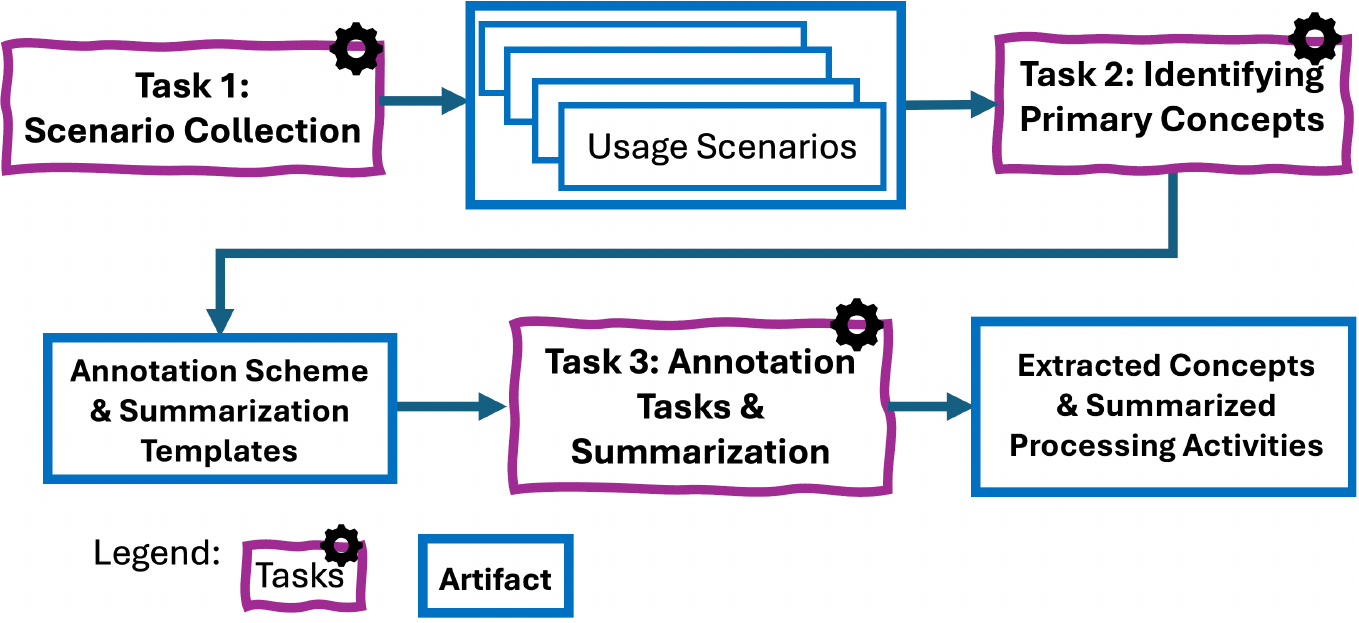}
   \centering
  \caption{The Overview of Extracting Processing Activities}
    \label{fig:approach}
    \vspace{-1em}
\end{figure}

We aim to generate segments of RoPA from usage scenarios provided by users as the primary data source entailing processing activities. The process of collecting scenarios and extracting processing activity summaries using extractive summarization is shown in Figure~\ref{fig:approach}. 

In Task 1, we first publish a survey to collect usage scenarios. In Task 2, given the usage scenarios as input, we identify the primary concepts that should be described in the processing activity summary. We employ a grounded analysis method to identify key concepts and design two \textit{artifacts} accordingly: (a) an annotation scheme; and (b) controlled-natural language templates to represent the summaries. In Task 3, we annotate scenarios using the annotation scheme and construct the summaries using the templates. 

% for collecting the required data. %Our approach to designing the studies is discussed in~\ref{sec:evaluation}. 

\subsection{Task 1: Scenario Collection} 

\begin{figure}
{%
\setlength{\fboxsep}{5pt}%
\setlength{\fboxrule}{0.5pt}%
  \fbox{\includegraphics[width=0.93\columnwidth, keepaspectratio]{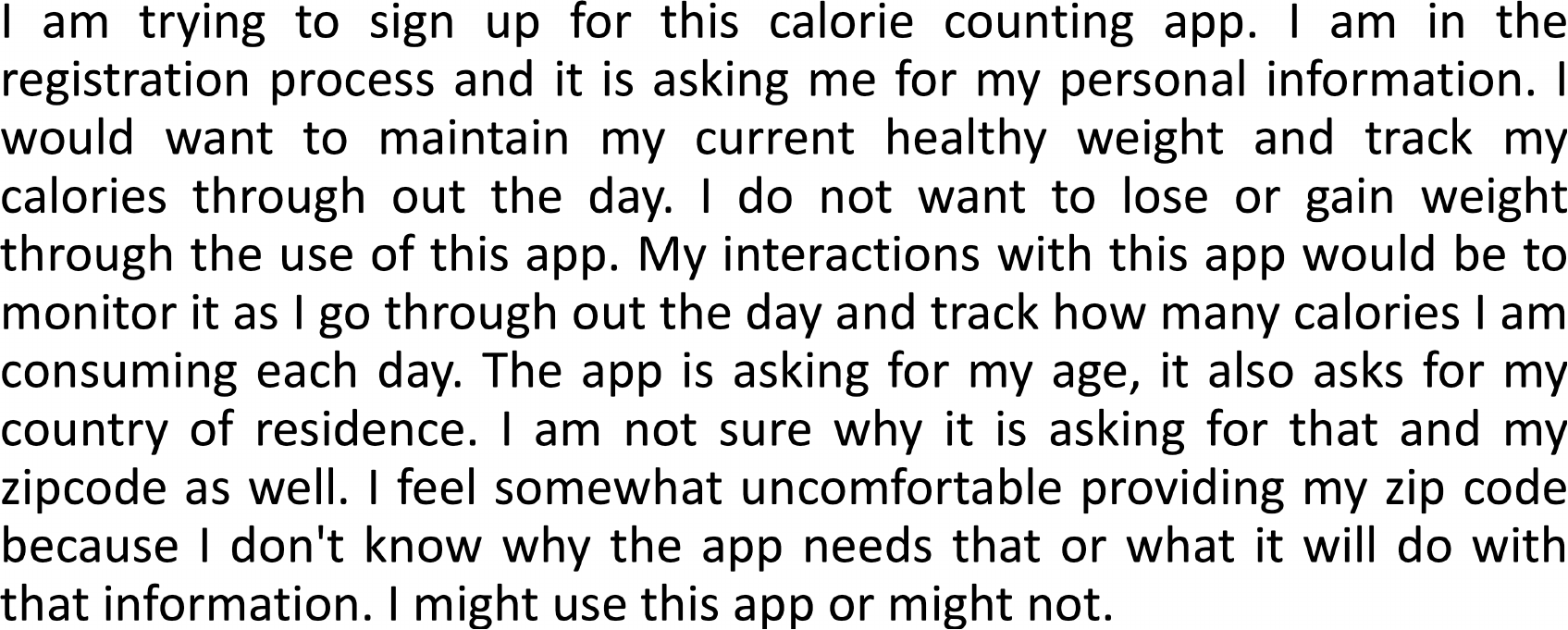}
  }%
  }%
   \centering
  \caption{Example Scenario}
    \label{fig:Example-Scenario}
    \vspace{-1em}
\end{figure}

%\mitra[inline]{Add why scenarios are better understood by stakeholders compared to software specifications}
To construct a corpus of usage scenarios, we follow the survey design method from Huang et al.~\cite{huang2023mobile}. To this end, we publish a survey and invite mobile app end-users to upload a screenshot from a mobile app, describing a usage scenario for the screenshot in at least 150 words, and answering some questions. The survey targets apps from various domains. 
%The study has been approved by the Institutional Review Board (IRB) and published through this link\footnote{https://galadriel.cs.utsa.edu/~mitrabokaei/scenario-study}.
Survey participants are first asked to select a mobile app they use frequently. Next, participants select a specific screen in the app and take a screenshot. Participants are instructed to select screens with the following properties: (a) emphasis on the core functionality of the apps; (b) not the app's homepage; (c) not the app's login page; and (d) not the app's settings page. We also provide the participants with examples of ideal and bad screenshots as further guidance in the instructions. To submit the screenshot from mobile phones, participants scan a QR code that navigates them to a web page where they upload, redact any personal information, and submit the screenshot. Fourth, the submitted screenshot is loaded to the main survey page, where participants are asked to write a usage scenario of at least 150 words that contains: (a) a description of the goal the user wants to achieve through the screen; (b) steps that they take to get to the screen on the app; and (c) the steps they take to achieve the goal once at the screen. The participants are also provided with an example of a scenario in the instructions.  
%\todo{add figure and link to the survey}
Figure~\ref{fig:Example-Scenario} illustrates an example of a scenario.

The survey was published on the AMT platform. Eligible participants were required to have completed at least 5,000 Human Intelligence Tasks (HITs), possess an approval rating exceeding 97\%, and reside in the US. Upon completion of the survey, workers received a compensation of \$4.00. The survey yielded a total of \textbf{50} usage scenarios.

%As part of our protocol to protect human subjects, workers must provide informed consent before participating in the survey and the study is monitored by our Institutional Review Board (IRB). 

%\begin{table}[ht!]
%\centering
%\caption{App Scenario Frequency for Apple \& Google App Category}
%\label{tab:appleGoogle_categories}
%\begin{tabular}{|l|c|c|}
%\hline
%\textbf{App Category} & \textbf{Apple} & \textbf{Google} \\
%\hline
%Education & 2 & -\\
%Finance &  7 & 1\\
%Food \& Drink &  1 & 4\\
%Health \& Fitness &  4 & 5\\
%Reference &  1 & -\\
%Shopping &  1 & 3\\
%Social Networking & 1  & 5\\
%Sports & 2 & 1\\
%Travel &  1 & 2\\
%Books \& Reference &  - & 1\\
%Casual & - & 2\\
%Entertainment & - & 1\\
%Lifestyle & - & 1\\
%Maps \& Navigation & - & 1\\
%Music \& Audio & - & 1\\
%Tools & - & 2\\
%\hline
%\textbf{Total} & \textbf{20} & \textbf{30} \\
%\hline
%\end{tabular}
%\vspace{0.2cm}
%\end{table}

%In Table~\ref{tab:appleGoogle_categories}, we present the frequencies of apps per category described by the Apple App and Google Play stores, for the 50 scenarios authored by users. The table shows the diversity of scenarios we collected from various domains and strengthens the generalizability of our work. 

\subsection{Task 2: Identifying Primary Concepts} \label{sec:approach-manual}
Usage scenarios may articulate actual or desired behaviors, indicating the mood of the scenario sentences~\cite{anton1998representational,jackson1995software, anton1994goal}. For instance, in the scenario depicted in Figure~\ref{fig:Example-Scenario}, the user expresses uncertainty about the app's need for a zip code. This expression conveys the user's desire for transparency and underscores the importance for app developers to incorporate a transparency requirement. Users' desires and preferences trigger change management \& maintenance activities. However, when constructing RoPA, our focus lies in identifying and extracting behaviors that represent the actual interactions between the system, users, and other entities. Thus, we propose an extractive summarization method tailored to filter processing activities from scenario texts, which may also include app deficiencies and desired features. 

To construct an extractive summarization method, we first need to identify the primary concepts involved in the actual behaviors stated in scenarios. 
For this reason, the first and the last authors conducted a manual grounded analysis of five randomly selected scenarios obtained in Task 1. Our grounded analysis yielded the following key concepts within the scenarios: (1) categories of actions performed; (2) actors (user, app, or external entities); (3) personal data types; (4) purposes of the actions; (5) recipients of data types (i.e., external entities); and (6) the User Interface (UI) elements that users interact with. Furthermore, we analyzed the action verbs present in the scenarios. As outlined in Task 1, users were instructed to provide the goal they aim to achieve, a list of steps they take to accomplish that goal, and data types used or provided during their interaction with the app (e.g., health data, financial data). This structural approach led to the identification of three distinct categories of actions within the scenarios:
%\mitra[inline]{provide examples for each verb}
\begin{itemize}
    \item \textit{Goal Actions}: Verbs or verb phrases that articulate the user's overarching goal to be achieved through the selected UI screen.
    \item \textit{Step Actions}: Verbs or verb phrases that detail the user's interaction with a specific UI component (e.g., button).
    \item \textit{Data Practice (DP) Actions}: Verbs or verb phrases that specify the events or actions related to the collection, usage, transfer, and retention of \textit{data types}.
\end{itemize}

Based on this analysis, we designed three controlled natural language (NL) templates that summarize sentences involving each category of actions. It is worth noting that a sentence in the scenario may encompass multiple action verbs from different categories. The use of these templates enables us to differentiate the concepts (e.g., data types, purposes, UI components, and actors) associated with each action verb and construct a controlled NL sentence that represents the action verb along with its corresponding concepts. This structured approach facilitates a concrete representation of the diverse actions and their associated concepts within the scenarios.

\begin{figure}[h]
    \includegraphics[width=1\linewidth]{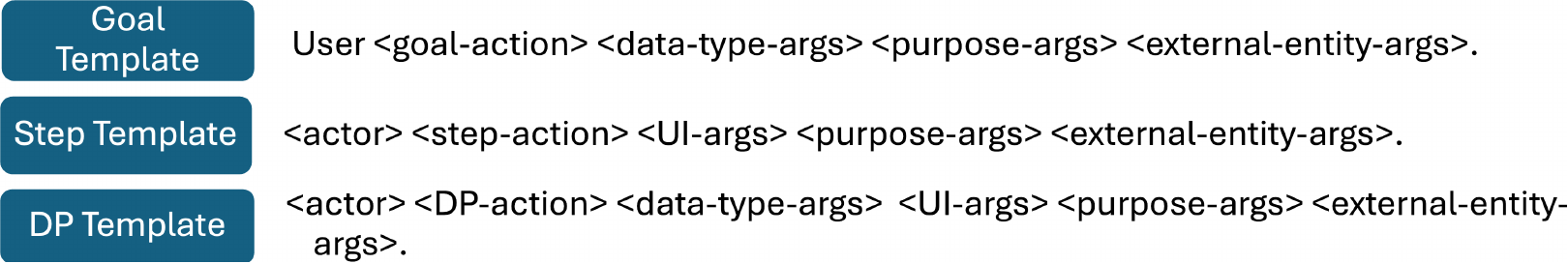}
    \caption{Controlled NL Templates for Summarization }
    \label{fig:template}
    %\vspace{-1em}
\end{figure}

Figure~\ref{fig:template} lists the templates for three categories of actions. For \textit{goal actions}, the template is composed of the token ``user'' that represents the actor, the goal action verb presented in a third person singular format, followed by three placeholders for data-type, purposes, and external-entity concept arguments (i.e., $\langle$data-type-args$\rangle$ $\langle$purpose-args$\rangle$ $\langle$external-entity-args$\rangle$). 
The template sentence for \textit{step actions} is composed of a placeholder token (i.e., $\langle$actor$\rangle$) for the actor, the step action verb presented in a third person singular format, followed by three placeholders for the UI component, purposes, and external-entity concepts (i.e., $\langle$UI-args$\rangle$ $\langle$purpose-args$\rangle$ $\langle$external-entity-args$\rangle$).
The template sentence for \textit{DP actions} is composed of a placeholder token (i.e., $\langle$actor$\rangle$) for the actor, the DP action verb presented in a third person singular format, followed by three placeholders for data-type, UI component, purposes, and external-entity concepts (i.e., $\langle$data-type-args$\rangle$ $\langle$UI-args$\rangle$ $\langle$purpose-args$\rangle$ $\langle$external-entity-args$\rangle$).

\begin{figure}[h]
    \includegraphics[width=1\linewidth]{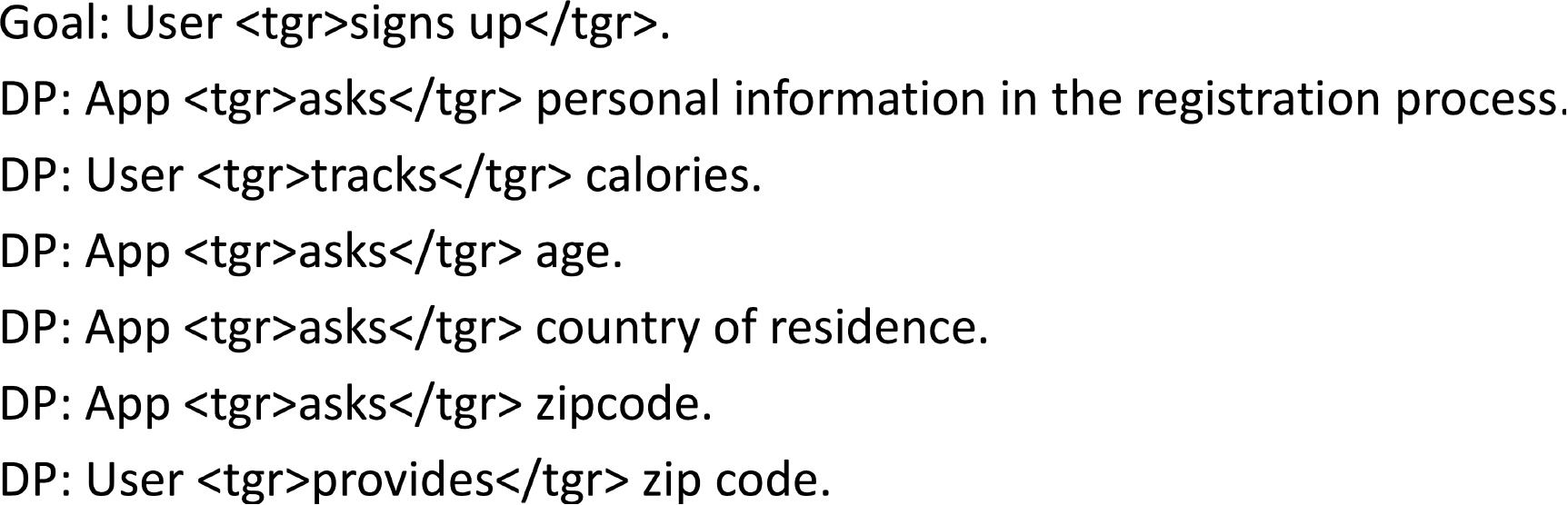}
    \caption{Filled Templates for the Example Scenario}
    \label{fig:filled-template}
\end{figure}

Figure~\ref{fig:filled-template} shows the summarized version of our example scenario, with filled templates for \textit{Goal} \& \textit{DP}, where the placeholder tokens are replaced by the arguments whenever possible. To specify the desired action for summarization, a special token $\langle$tgr$\rangle$ is employed surrounding the action verbs. If there are multiple arguments for the same slot (e.g., $\langle$data-type-args$\rangle$), we connect the arguments with the token ``and''.

\subsection{Task 3: Annotation \& Summarization}

\begin{figure}[h]
    \includegraphics[width=1\linewidth]{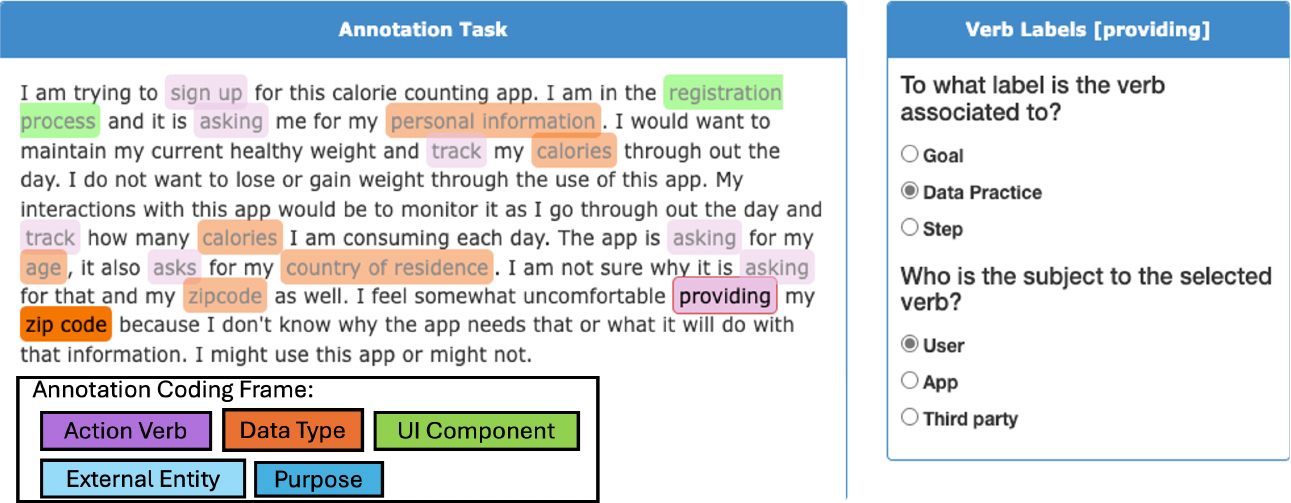}
    \caption{Annotation Coding Frame}
    \label{fig:annotation-scheme}
    \vspace{-1em}
\end{figure}

To generate processing activity summaries for all 50 scenarios, we developed an annotation tool and published it on AMT. The coding frame is illustrated in Figure~\ref{fig:annotation-scheme} and is based on the primary concepts identified in Task 2. This coding frame contains the following codes: \textit{action verb or verb phrase}, \textit{data type}, \textit{purposes of the actions}, \textit{external entities as recipients of data types}, and \textit{UI component}. The annotators are instructed to first highlight an action verb in a sentence and then identify two attributes for the action: the action category (i.e., goal, step, or data practice (DP)) and the actor (i.e., user, app, or external entity). Next, the annotators are instructed to identify the corresponding concepts for the annotated action verb, such as data types, purposes, external entities, and UI components. 

Among 50 scenarios, we generated a HIT per scenario. Two annotators (the first \& last authors) were assigned to the HITs such that each HIT would be annotated by two respondents. An annotator typically took about 9.65 minutes to complete one HIT. To facilitate reconciliation between the two annotators, we use tokenization and present each scenario as a sequence of tokens that are annotated with specific labels. %For this reason, we create a table where the first column represents the sequence of tokens from a scenario, and the second \& third columns represent each annotator's annotations for the token.

We annotated the 50 scenarios in four rounds. In round one, both authors coded a random sample of 10 scenarios using the coding frame, which yielded in Cohen Kappa of 0.50~\cite{cohen1960coefficient}, which is a moderate agreement~\cite{landis1977measurement}. The authors next discussed disagreements and identified heuristics to clarify boundary and edge cases. In round two, the authors coded a new random sample of 10 new scenarios using the new heuristics, yielding in Cohen Kappa of 0.68. The authors reconvened, examined disagreements, and developed the following heuristics: 

\begin{itemize}
    \item H1: Step action verbs involve UI components rather than specifying data types. 
    \item H2: Goal action verbs are introduced by a verb phrase, such as ``I want to$\cdots$'' or ``The app allows me to$\cdots$'', indicating a clear goal.
    \item H3: Users may describe app-wide steps or data practices rather than specific screen-related actions. Therefore, those action verbs cannot be considered as pre/post-conditions of the main goal for the scenario. 
    \item H4: Extra information detailing data types, purposes, and external entities (e.g., pronouns and articles) is not annotated. Only the core/head of a noun phrase is selected, excluding additional modifiers.
    \item H5: Scenarios with lists of elements separated by commas, conjunctions/disjunctions are annotated individually. 
\end{itemize}  

In round three, the authors re-coded the second sample using the heuristics to reach a Kappa of 0.74. In round 4, the authors coded a new random sample of 10 scenarios using the heuristics and coding frame to yield a Kappa of 0.95 (almost perfect agreement)~\cite{landis1977measurement}. Due to this high level of agreement, the remaining scenarios were then coded by the last author using the coding frame and accompanying heuristics. To this point, we have a total of 400 action verbs from our reconciliation of 50 annotated scenarios. Of these, 64 are linked to goal, 83 to step, and 253 to DP.
% 6 instances of anaphora and 32 instance where we had to update the template to add inference

\section{Experiment Design}\label{sec:experimentDesign}
We investigate how LLMs can extract summaries of processing activities from scenarios automatically. We explore the effects of the number of training examples in few-shot learning, repetition, and ordering of the examples on LLMs performance for extractive summarization. From the 50 scenarios in the corpus, we identify sentences that contain goal, step, or DP action verbs, yielding three datasets containing 64, 83, and 253 sentences, respectively. 
We partition each dataset into training, validation, and testing sets using a ratio of 60:20:20. 
Using the training, validation, and testing sets for all three datasets (i.e., goal, step, and DP), we design two experiments to address the following research questions. 

\noindent \textbf{RQ1.} How can the number of examples in few-shot learning change the performance of extractive summarization?

\noindent \textbf{RQ2.} How consistent are LLMs when repeating the same experiment in terms of providing identical results? 

\noindent \textbf{RQ3.} How crucial is the order sensitivity in few-shot learning?

\noindent \textbf{RQ4.} To what extent can LLMs effectively extract RoPA concepts and construct summaries from usage scenarios?

To design our experiments, we utilize GPT-3.5 Turbo as an instance of LLMs. 
We design Experiment 1 to investigate the effect of the number of examples in few-shot learning on the validation set. We further evaluate the effect of repetition on the results generated by GPT-3.5 Turbo. 
Building on the optimal number of examples identified in Experiment 1, we formulate Experiment 2 to investigate the order \& arrangement of examples within the prompt. Using the results from both experiments, we configure a few-shot learning model and evaluate it on the testing set for each dataset.

\subsection{Experiment 1} \label{sec:experiment1}

\begin{figure}
  \includegraphics[width=\columnwidth, keepaspectratio]{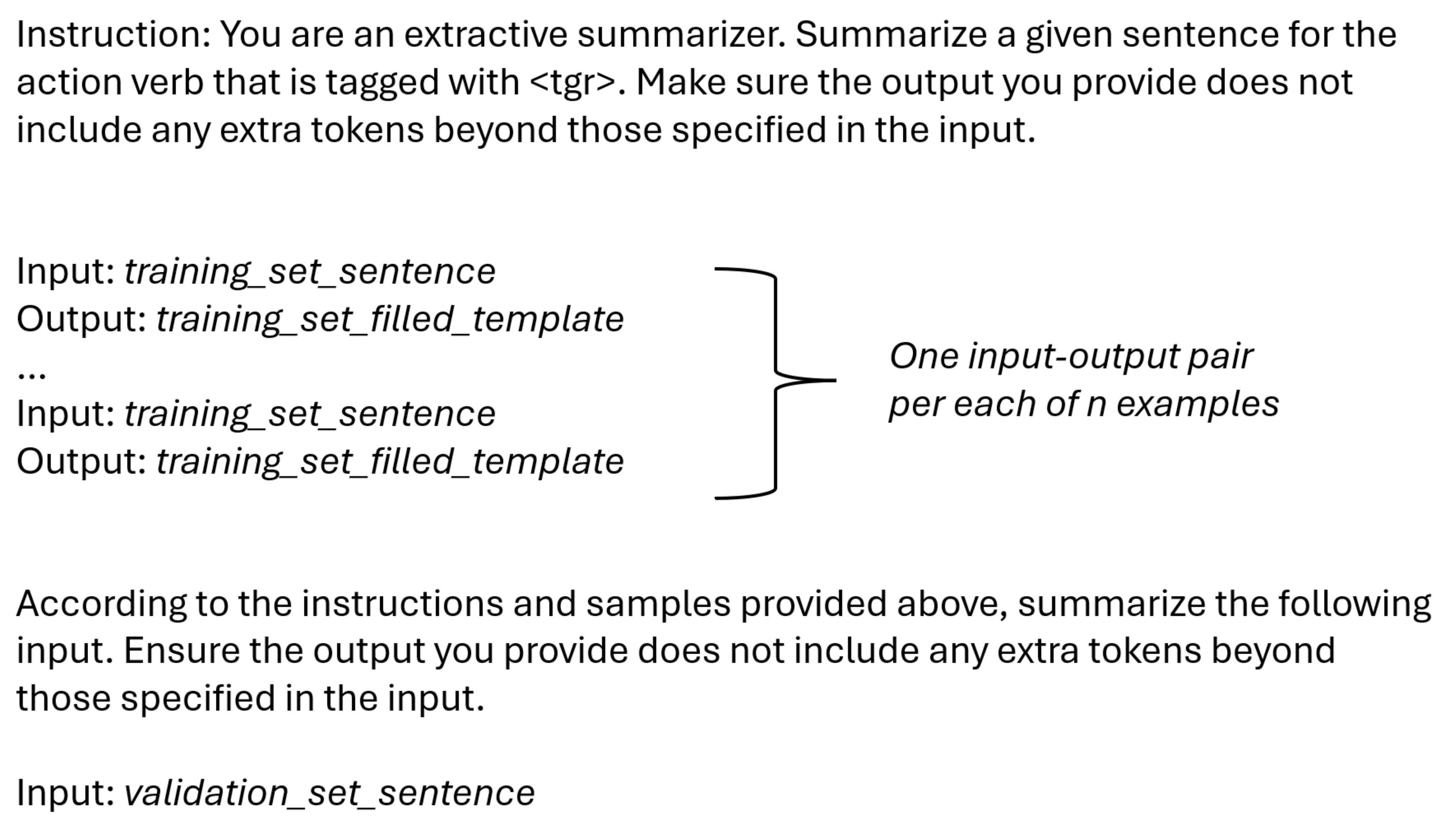}
   \centering
  \caption{Prompt Structure for Zero to 10 Shot Examples}
    \label{fig:prompt}
    \vspace{-1em}
\end{figure}

We aim to explore the significance of the number of examples in few-shot learning prompts and the resulting variation when the experiment is repeated. For each dataset, we initiate this experiment by randomly selecting 10 examples from the training set. %, employing a fixed seed value of 23 to ensure reproducibility. 
Following this, we structure prompts in a specific format to guide the Chat Completion API of GPT-3.5 Turbo as shown in Figure~\ref{fig:prompt}. 
This format allows us to change the number of examples from zero to 10 to create \textit{11 unique prompts}. Additionally, to gauge response consistency, each prompt was repeated 10 times (10 different GPT API calls). 
Note that we consider the zero-shot prompt as the \textit{baseline} for this experiment. 
The prompts begin with an ``instruction'' token, followed by specifying a persona for GPT as shown in Figure~\ref{fig:prompt}. Research suggests that including a persona in prompt instructions can enhance the performance of LLMs ~\cite{chen2023unleashing},~\cite{yu2023exploring}, ~\cite{oleaevaluating},~\cite{white2023prompt}. Subsequently, the instructions detail the summarization task, concluding with a constraint to ensure that the summary output only contains tokens from the input. Examples are then provided based on the number of shots used in the experiments. Additionally, an excerpt is provided for GPT to reference the initial instruction in generating an output for the given input~\cite{white2023prompt},~\cite{chen2023unleashing}. It's worth noting that altering and assessing this prompt's structure can offer insights into its impact on results. However, in this paper, we narrow our focus to only exploring how variations in the number of examples and the order of examples influence the result.

Evaluation is conducted using six different metrics, including ROUGE-1, ROGUE-2, ROUGE-L, ROGUE-S, METEOR, and BERTScore, on the testing set. ROUGE-1 focuses on individual word overlap, ROUGE-2 considers adjacent word pairs, while ROUGE-L assesses the longest common subsequence, allowing for word reordering \cite{akter2022revisiting}. ROUGE-S incorporates skip-bigrams, introducing flexibility in word order for sentence-level coherence \cite{nanba2006automatic}. METEOR extends evaluation beyond exact word matches, considering synonymy and stemming for a more nuanced semantic assessment \cite{banerjee2005meteor}. Lastly, BERTScore captures the detailed semantic similarities between the provided reference and prediction\cite{soleimani2023nonfacts}.

%This meticulous experiment design seeks to unravel the impact of the number of shots (i.e., examples) in prompts, offering insights into the performance variation and consistency of GPT-3.5 Turbo across multiple repetitions.

\subsection{Experiment 2}
%To address RQ3 regarding the sensitivity of order in few-shot learning, we design Experiment 2. In Experiment 2, we evaluate whether a specific order of samples can be achieved to get the optimal performance in few-shot learning. For this evaluation, we find all the permutation of the examples that we obtain from Experiment 1, and run the prompt on a validation set. 

%We design Experiment 2 to address RQ3 regarding the sensitivity of order in few-shot learning. 
Our objective is to determine if organizing examples in a specific order in prompts improves the overall effectiveness of few-shot learning models. To achieve this, we explore all order permutations of examples (i.e., number of shots) derived from Experiment 1 for each dataset (i.e., goal, step, and DP). Using the order permutations and the number of examples derived from Experiment 1, we create unique prompts and evaluate each prompt on the validation sets of each dataset. Similar to Experiment 1, we use GPT-3.5 Turbo and evaluate each prompt on the validation sets. The prompts in this experiment have a similar structure as Figure~\ref{fig:prompt}. 

%To accomplish this, we establish a ``user'' role using instructions for the model, a permutation of examples, and an instance-marked sentence for which a response is required, as depicted in Figure X. 

%We pick ROUGE-L score since it looks for longest in-sequence common n-grams which thereby eliminating the necessity of defining an n value for the n-gram match .For each permutation, we calculate the average F1 ROUGE-L score obtained when tested on the validation set. Subsequently, we compute the variance of the F1 ROUGE-L scores across all permutations to see sensitivity to order in few-shot learning.

\section{Evaluation \& Results}\label{sec:results}

%for the list of evaluation metrics refer to "A survey of evaluation metrics used for NLG systems" paper~\cite{sai2022survey}.

% Results For Step - Random 9 example Rogue-1: 0.7582352941176471 Rogue-2: 0.6258823529411766 Rogue-L: 0.7582352941176471 Rogue-S: 0.4870588235294118 METEOR: 0.8488235294117646 BERTScore: 0.9547058823529412

% Results For Step - Random 6 example Rogue-1: 0.6249019607843137 Rogue-2: 0.4209803921568628 Rogue-L: 0.6113725490196078 Rogue-S: 0.3037254901960784 METEOR: 0.6929411764705882 BERTScore: 0.9392156862745098

% Results For Goal - Random 7 example Rogue-1: 0.75 Rogue-2: 0.4753846153846154 Rogue-L: 0.75 Rogue-S: 0.36230769230769233 METEOR: 0.8038461538461538 BERTScore: 0.9538461538461538
% Table X, Y, and Z shows the average F1 scores over 6 metrics for n shot ranging from 0 to 10 for Goal, Step, and Data Practice respectively. Table X, Y, and Z also presents the variance for each of the shots.  

\begin{table*}[h]
\centering
\caption{Performance Metrics for Different Number of Examples}
\begin{tabular}{|l|lll|lll|lll|lll|lll|lll|}
\hline
            & \multicolumn{3}{c|}{ROUGE-1}                               & 
            \multicolumn{3}{c|}{ROUGE-2}                               &
            \multicolumn{3}{c|}{ROUGE-L}                               &
            \multicolumn{3}{c|}{ROUGE-S}                               &
            \multicolumn{3}{c|}{METEOR}                              &
            \multicolumn{3}{c|}{BERTScore}                               \\
            \hline
Shots & \multicolumn{1}{c|}{Goal} & \multicolumn{1}{c|}{Step} & DP &
\multicolumn{1}{c|}{Goal} & \multicolumn{1}{c|}{Step} & DP & 
\multicolumn{1}{c|}{Goal} & \multicolumn{1}{c|}{Step} & DP &
\multicolumn{1}{c|}{Goal} & \multicolumn{1}{c|}{Step} & DP & 
\multicolumn{1}{c|}{Goal} & \multicolumn{1}{c|}{Step} & DP &
\multicolumn{1}{c|}{Goal} & \multicolumn{1}{c|}{Step} & DP \\ \hline

0           & \multicolumn{1}{c|}{0.26}    & \multicolumn{1}{c|}{0.22}    & 0.22  & \multicolumn{1}{c|}{0.10}    & \multicolumn{1}{c|}{0.12}    & 0.10 &  
\multicolumn{1}{c|}{0.25}    & \multicolumn{1}{c|}{0.21}    & 0.22  & 
\multicolumn{1}{c|}{0.09}    & \multicolumn{1}{c|}{0.11}    & 0.09  & 
\multicolumn{1}{c|}{0.4}    & \multicolumn{1}{c|}{0.45}    & 0.31  & 
\multicolumn{1}{c|}{0.89}    & \multicolumn{1}{c|}{0.89}    & 0.88  \\ \hline

1           & \multicolumn{1}{c|}{0.62}    & \multicolumn{1}{c|}{0.62}    & 0.53  & \multicolumn{1}{c|}{0.47}    & \multicolumn{1}{c|}{0.44}    & 0.35  &
\multicolumn{1}{c|}{0.62}    & \multicolumn{1}{c|}{0.61}    & 0.53  &
\multicolumn{1}{c|}{0.43}    & \multicolumn{1}{c|}{0.42}    & 0.32  &
\multicolumn{1}{c|}{0.68}    & \multicolumn{1}{c|}{0.75}    & 0.66  &
\multicolumn{1}{c|}{0.95}    & \multicolumn{1}{c|}{0.95}    & 0.94 \\ \hline

2           & \multicolumn{1}{c|}{0.63}    & \multicolumn{1}{c|}{0.68}    & 0.54  & \multicolumn{1}{c|}{0.49}    & \multicolumn{1}{c|}{0.58}    & 0.36  &
\multicolumn{1}{c|}{0.63}    & \multicolumn{1}{c|}{0.68}    & 0.54  &
\multicolumn{1}{c|}{0.45}    & \multicolumn{1}{c|}{0.54}    & 0.32  &
\multicolumn{1}{c|}{0.68}    & \multicolumn{1}{c|}{0.79}    & 0.63  &
\multicolumn{1}{c|}{0.95}    & \multicolumn{1}{c|}{0.96}    & 0.94  \\ \hline

3           & \multicolumn{1}{c|}{0.63}    & \multicolumn{1}{c|}{0,71}    & 0.58  & \multicolumn{1}{c|}{0.47}    & \multicolumn{1}{c|}{0.60}    & 0.41  &
\multicolumn{1}{c|}{0.63}    & \multicolumn{1}{c|}{0.71}    & 0.57  &
\multicolumn{1}{c|}{0.43}    & \multicolumn{1}{c|}{0.58}    & 0.38  &
\multicolumn{1}{c|}{0.68}    & \multicolumn{1}{c|}{0.79}    & 0.68  &
\multicolumn{1}{c|}{0.95}    & \multicolumn{1}{c|}{0.96}    & 0.95  \\ \hline

4           & \multicolumn{1}{c|}{0.65}    & \multicolumn{1}{c|}{0.72}    & 0.56  & \multicolumn{1}{c|}{0.47}    & \multicolumn{1}{c|}{0.63}    & 0.37  &
\multicolumn{1}{c|}{0.64}    & \multicolumn{1}{c|}{0.72}    & 0.55  &
\multicolumn{1}{c|}{0.43}    & \multicolumn{1}{c|}{0.59}    & 0.34  &
\multicolumn{1}{c|}{0.69}    & \multicolumn{1}{c|}{0.79}    & 0.67  &
\multicolumn{1}{c|}{0.95}    & \multicolumn{1}{c|}{0.96}    & 0.94  \\ \hline

5           & \multicolumn{1}{c|}{0.67}    & \multicolumn{1}{c|}{0.71}    & 0.61  & \multicolumn{1}{c|}{0.49}    & \multicolumn{1}{c|}{0.63}    & 0.45  &
\multicolumn{1}{c|}{0.67}    & \multicolumn{1}{c|}{0.71}    & 0.60  &
\multicolumn{1}{c|}{0.45}    & \multicolumn{1}{c|}{0.59}    & 0.41  &
\multicolumn{1}{c|}{0.73}    & \multicolumn{1}{c|}{0.78}    & 0.70  &
\multicolumn{1}{c|}{0.95}    & \multicolumn{1}{c|}{0.96}    & 0.95  \\ \hline

6           & \multicolumn{1}{c|}{0.71}    & \multicolumn{1}{c|}{0.72}    & 0.63  & \multicolumn{1}{c|}{0.53}    & \multicolumn{1}{c|}{0.64}    & 0.48  &
\multicolumn{1}{c|}{0.71}    & \multicolumn{1}{c|}{0.72}    & 0.63  &
\multicolumn{1}{c|}{0.48}    & \multicolumn{1}{c|}{0.60}    & 0.45  &
\multicolumn{1}{c|}{0.77}    & \multicolumn{1}{c|}{0.79}    & 0.74  &
\multicolumn{1}{c|}{0.96}    & \multicolumn{1}{c|}{0.96}    & 0.95  \\ \hline

7           & \multicolumn{1}{c|}{0.71}    & \multicolumn{1}{c|}{0.72}    & 0.64  & \multicolumn{1}{c|}{0.51}    & \multicolumn{1}{c|}{0.63}    & 0.49  &
\multicolumn{1}{c|}{0.71}    & \multicolumn{1}{c|}{0.72}    & 0.63  &
\multicolumn{1}{c|}{0.46}    & \multicolumn{1}{c|}{0.61}    & 0.45  &
\multicolumn{1}{c|}{0.75}    & \multicolumn{1}{c|}{0.79}    & 0.73  &
\multicolumn{1}{c|}{0.96}    & \multicolumn{1}{c|}{0.96}    & 0.95  \\ \hline

8           & \multicolumn{1}{c|}{0.69}    & \multicolumn{1}{c|}{0.71}    & 0.63  & \multicolumn{1}{c|}{0.51}    & \multicolumn{1}{c|}{0.62}    & 0.47  &
\multicolumn{1}{c|}{0.69}    & \multicolumn{1}{c|}{0.71}    & 0.63  &
\multicolumn{1}{c|}{0.47}    & \multicolumn{1}{c|}{0.59}    & 0.44  &
\multicolumn{1}{c|}{0.73}    & \multicolumn{1}{c|}{0.80}    & 0.73  &
\multicolumn{1}{c|}{0.96}    & \multicolumn{1}{c|}{0.96}    & 0.95  \\ \hline

9           & \multicolumn{1}{c|}{0.70}    & \multicolumn{1}{c|}{0.70}    & 0.62  & \multicolumn{1}{c|}{0.52}    & \multicolumn{1}{c|}{0.61}    & 0.47  &
\multicolumn{1}{c|}{0.70}    & \multicolumn{1}{c|}{0.70}    & 0.62  &
\multicolumn{1}{c|}{0.48}    & \multicolumn{1}{c|}{0.58}    & 0.44  &
\multicolumn{1}{c|}{0.72}    & \multicolumn{1}{c|}{0.78}    & 0.71  &
\multicolumn{1}{c|}{0.96}    & \multicolumn{1}{c|}{0.96}    & 0.95  \\ \hline

10           & \multicolumn{1}{c|}{0.74}    & \multicolumn{1}{c|}{0.70}    & 0.64  & \multicolumn{1}{c|}{0.58}    & \multicolumn{1}{c|}{0.61}    & 0.48  &
\multicolumn{1}{c|}{0.74}    & \multicolumn{1}{c|}{0.70}    & 0.63  &
\multicolumn{1}{c|}{0.54}    & \multicolumn{1}{c|}{0.57}    & 0.44  &
\multicolumn{1}{c|}{0.75}    & \multicolumn{1}{c|}{0.77}    & 0.73  &
\multicolumn{1}{c|}{0.96}    & \multicolumn{1}{c|}{0.96}    & 0.95  \\ \hline

\end{tabular}
\label{tab:performanceMetricsRQ2}
\end{table*}

% \begin{figure*}
% \centering  
% \subfigure{\includegraphics[width=0.32\textwidth]{figures/GoalRepetition1.pdf}} 
%   \subfigure{\includegraphics[width=0.32\textwidth]{figures/StepRepetition1.pdf}} 
%   \subfigure{\includegraphics[width=0.32\textwidth]{figures/DPRepetition1.pdf}}
% \caption{Repetition for Goal, Step, and DP}
% \label{fig:box-plots}
% \vspace{-1em}
% \end{figure*}

\begin{figure*}
\centering  

\includegraphics[width=0.32\textwidth]{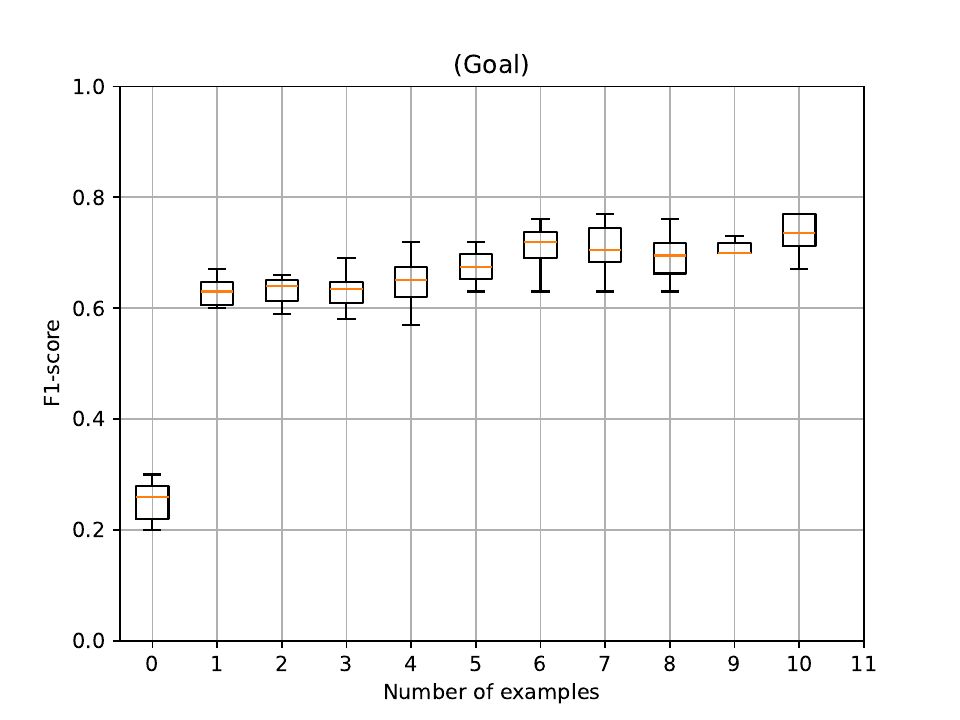}
\includegraphics[width=0.32\textwidth]{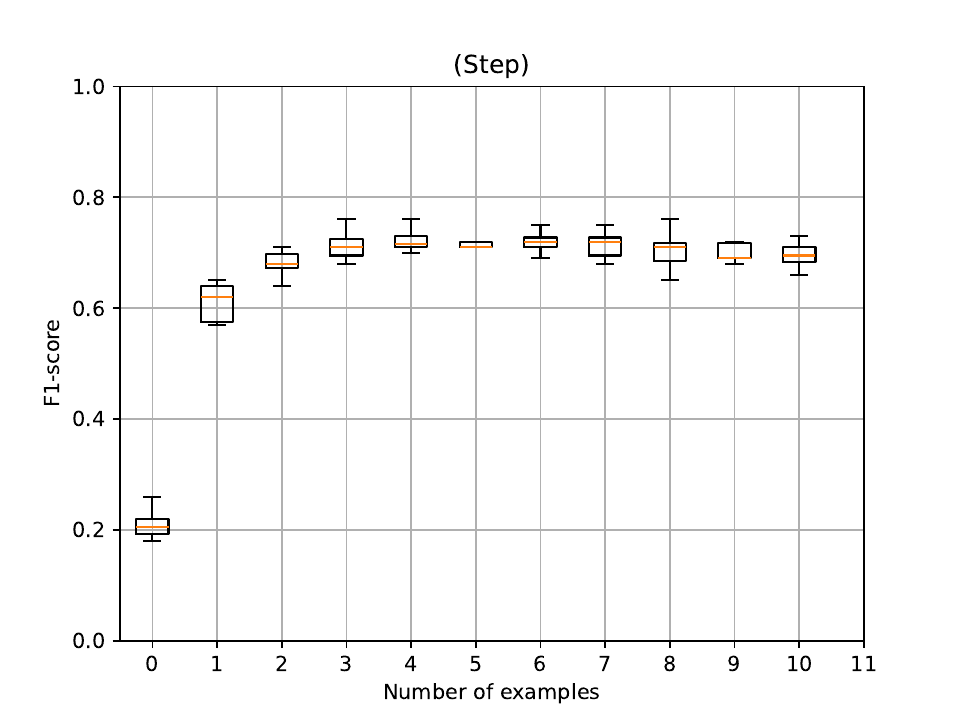}
\includegraphics[width=0.32\textwidth]{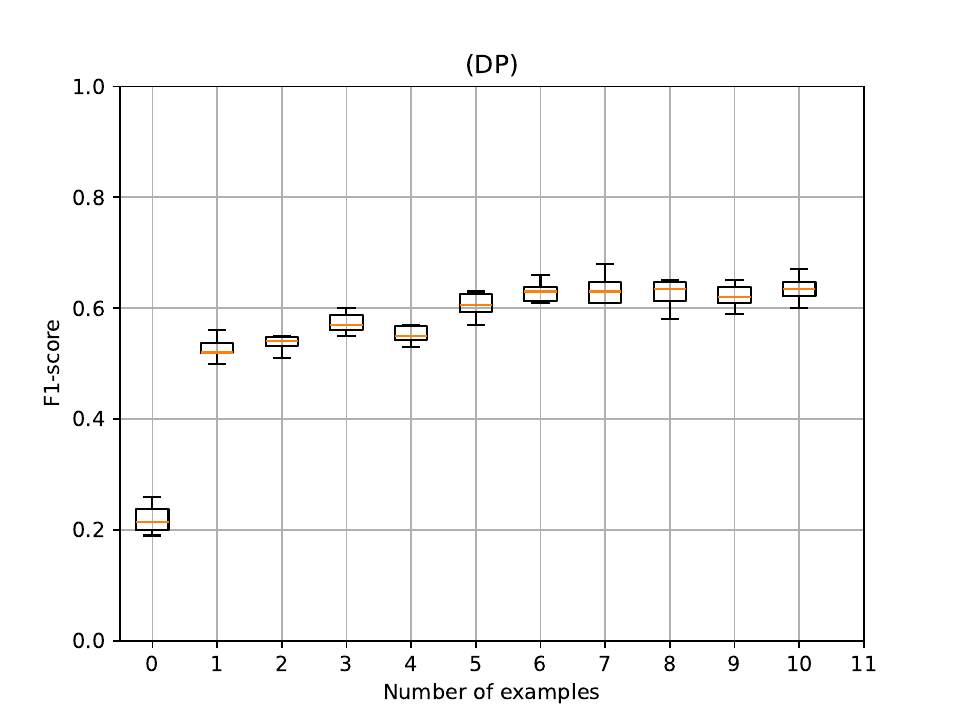}

\caption{Repetition for Goal, Step, and DP}
\label{fig:box-plots}
\vspace{-1em}
\end{figure*}

\begin{figure}
  \includegraphics[width=\columnwidth, keepaspectratio]{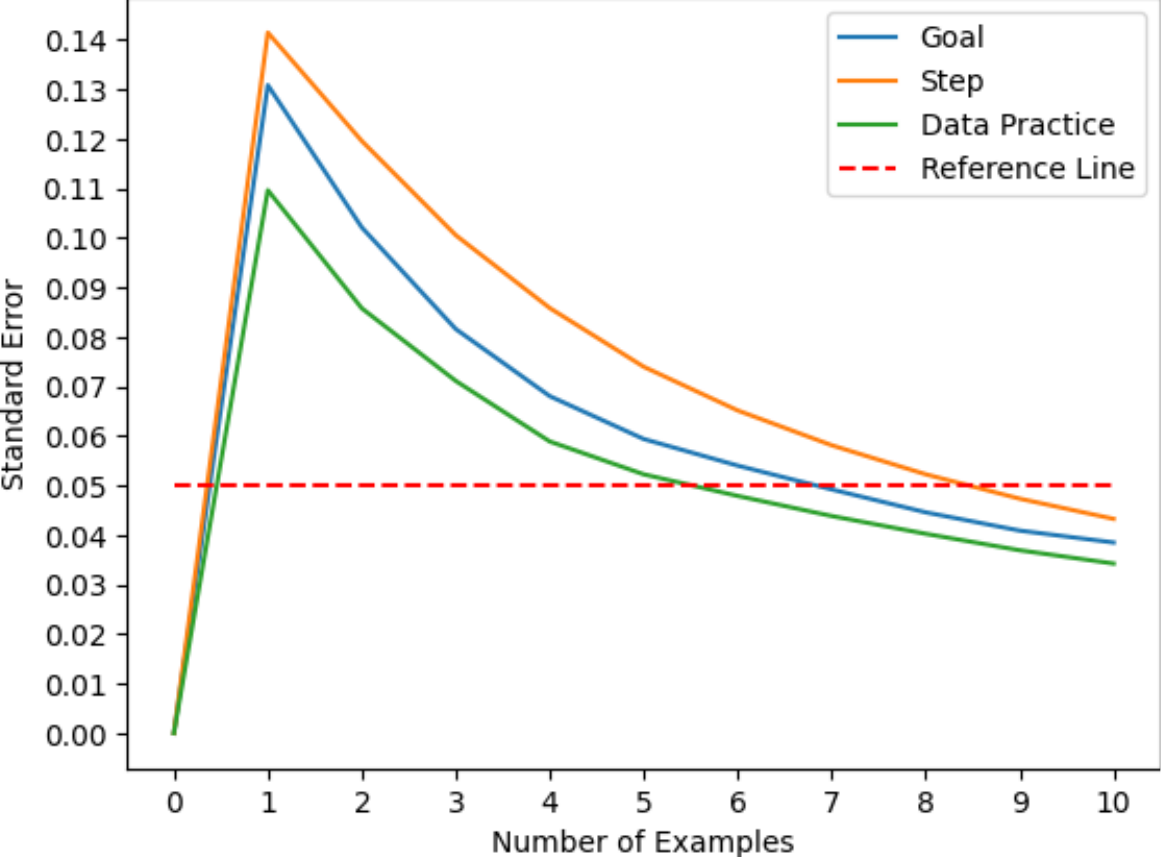}
   \centering
  \caption{Standard Error Line Plot}
    \label{fig:std}
    \vspace{-1em}
\end{figure}

\noindent\textbf{Experiment 1 Results:} 
Table~\ref{tab:performanceMetricsRQ2} displays the mean F1-scores for six evaluation metrics across different numbers of examples (i.e., zero to 10). The means are computed based on 10 repetitions of each unique prompt. Across all three datasets, there is a noticeable improvement in performance from zero-shot (i.e., baseline) to one-shot-learning for all metrics. ROUGE-L is a more restrictive metric that prioritizes the recall of content units shared between the input text and the output summary. It highlights the significance of capturing the reference text's overall content and meaning. Therefore, we opt for the ROUGE-L score for detailed analysis.  
Figure~\ref{fig:box-plots} illustrates the box plots representing the mean Rouge-L F1 scores for the number of examples changing from zero to 10, along with the variance for each shot. Upon examination to address \textbf{RQ1}, we note that the variance remains insignificant for all repetitions of each unique prompt. 

To address \textbf{RQ2} and determine the optimal number of examples for few-shot learning, we calculate the cumulative mean of Rouge-L F1 score for zero to 10 shots. Additionally, we calculate the standard error for each cumulative mean. We expect that the standard error decreases as the number of examples increases. Figure~\ref{fig:std} presents a line plot depicting the relationship between the standard error and the number of examples. As expected, an increase in the number of examples results in a decrease in the standard error. To promote reproducibility, we establish an acceptable standard error threshold of 0.05. This threshold may vary depending on researchers' available resources to augment the number of examples in few-shot learning. The dashed reference line in Figure~\ref{fig:std} represents this 0.05 acceptable error threshold. Considering this level of error, we select seven, nine, and six examples for the goal, step, and DP datasets. Our results can be found online%~\footnote{http://tinyurl.com/29j499v8}.
~\footnote{https://tinyurl.com/DataForESPRE2024}.

\begin{figure}
  \includegraphics[width=\columnwidth, keepaspectratio]{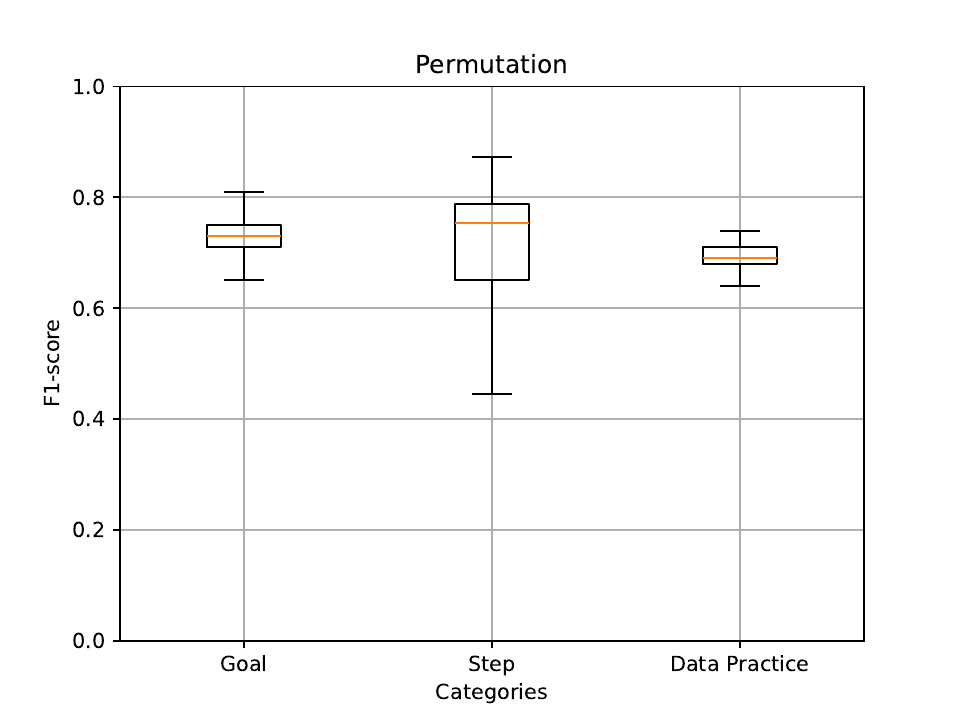}
   \centering
  \caption{Permutations Performed on Each Dataset}
    \label{fig:permutation}
    \vspace{-1em}
\end{figure}

\noindent\textbf{Experiment 2 Results:} Figure~\ref{fig:permutation} illustrates the distribution of ROUGE-L F1 scores for the order permutations of 7!, 9!, and 6! for the goal, step, and DP validation sets. Each box plot illustrates minimal variability in the distribution of F1 scores, as indicated by the size of the box and whiskers. To address \textbf{RQ3}, we report the variance for the goal, step, and DP validation sets as 0.00, 0.06, and 0.00, respectively.

To optimize the permutation process and minimize execution time, we leverage the ``ThreadPoolExecutor'' library for efficient task management across multiple threads concurrently. This strategy ensures that tasks are assigned to idle threads as they are completed, maximizing parallelism and speeding up the overall process. Additionally, we distribute the permutation workload across multiple files, enabling parallel execution within each file and across different files simultaneously. We utilized a high-performance virtual machine equipped with 80 CPU cores and 128GB of memory, further enhancing computational capabilities. 
The execution time for goal and DP permutations was approximately 24 hours, while for the step validation set, it extended slightly beyond 48 hours. The cumulative cost of completing these API calls for the permutations amounted to roughly \$3,500.

\noindent\textbf{Testing Set Results:} 
%We leverage the findings from Experiments 1 \& 2 to construct three prompts tailored to the goal, step, and DP datasets. 
Experiment 1 \& 2 results reveal that the number of examples within a prompt significantly influences performance, while repetition and order permutations have a minor impact on consistency \& performance. Consequently, we leverage these findings and use one prompt per dataset, comprising seven, nine, and six examples for the goal, step, and DP datasets, respectively. We assess each test set using these prompts and GPT-3.5 Turbo to address \textbf{RQ4}.
Table~\ref{tab:finalResultsOnTestingSet} presents the results for the testing sets.

\vspace{-1em}
\begin{table}[ht!]
\caption{Results on Testing Set}
\resizebox{1\columnwidth}{!}{
\begin{tabular}{|l|l|l|l|l|l|l|}
\hline
Metric        & ROUGE-1 & ROUGE-2 & ROUGE-L & ROGUE-S & METEOR & BERTScore \\ \hline
Goal          & 0.75       & 0.48       & 0.75       & 0.36       & 0.80      & 0.95         \\ \hline
Step          & 0.75       & 0.63       & 0.76       & 0.49       & 0.85      & 0.95         \\ \hline
Data Practice & 0.62       & 0.42       & 0.61       & 0.30       & 0.69      & 0.94         \\ \hline
\end{tabular}}
\label{tab:finalResultsOnTestingSet}
\vspace{-1em}
\end{table}

%We will use human judgments, ROUGE, METEOR, and BERTScore

\section{Discussion}\label{sec:discussion}

We proceed to examine the research questions in light of our findings. 
\textbf{RQ1} delves into how varying the number of examples in few-shot learning impacts the performance of extractive summarization. We ascertain the optimal number of shots required by analyzing the accumulative mean and standard error of ROUGE-L scores. %For our experiment, we establish the standard error threshold as 0.05. %Utilizing this threshold as an acceptable error, we observe that the line plots for each dataset consistently fall below the threshold, indicating optimal example shots of nearly seven, nine, and six for the goal, step, and DP datasets, respectively. 
Further, we set zero-shot learning as our baseline for Experiment 1. We observe that zero-shot learning performs poorly for this specific task across all datasets.  Therefore, it is crucial to provide examples for such domain-specific tasks. 

\textbf{RQ2} evaluates the reliability of GPT-3.5 Turbo by examining its consistency in producing identical results when the same experiment is repeated. Through repetition of each shot 10 times, our analysis presents a minimal variance, approaching zero. This observation underscores the consistency exhibited by GPT-3.5 Turbo across repeated iterations, affirming its ability to generate consistent outputs. 

\textbf{RQ3} investigates the sensitivity of order in few-shot learning through permutation analysis. Using optimal example counts of seven, nine, and six for goal, step, and DP, respectively, we generate permutations of 7!, 9!, and 6! for each dataset. Subsequently, these permutations are evaluated on a validation set. Upon comparing the ROUGE-L F1 scores across all permutations, we observe minimal variance. This leads us to conclude that the order of examples in GPT-3.5 Turbo has a negligible impact on few-shot learning outcomes.

To address \textbf{RQ4} and evaluate the effectiveness of GPT-3.5 Turbo in extracting RoPA concepts and generating summaries, we employ the optimal number of examples to craft three distinct prompts, which are then applied to the testing sets. Our prompts achieve successful summarization of processing activities, yielding an average ROUGE-L F1 score of 70\%. 
To further understand the metric scores, we expand our exploration with a manual qualitative study. 
In this study, both the first and last authors compared the ground truth summary templates with the summaries generated by GPT-3.5 Turbo for each sentence in the three testing sets. This comparison resulted in the development of a coding scheme comprising six code categories: (1) additional modifiers and adjectives; (2) incorrect action verb or subject; (3) missing data type; (4) missing purpose; (5) missing UI Component; and (6) summary contains more than two verbs. After completing the coding exercise, both annotators reconciled any disagreements and analyzed the results. From the analysis of 81 sentences in the testing sets across all three datasets, the following ratios were observed for the coding categories 1-6: 29/81, 20/81, 2/81, 6/81, 1/81, and 6/81. 
Coding category (1) exhibits the highest ratio, indicating instances where GPT-3.5 Turbo extracted additional tokens from the sentence compared to human extractions in the ground truth. 
For example, GPT-3.5 Turbo generates ``User gets regular promotions offered,'' for a testing set sentence, ``If I opt in, I would probably be able to $\langle$tgr$\rangle$get$\langle$/tgr$\rangle$ regular promotions offered to me,'' with the ground truth summary template as ``User gets promotions''. Despite these disparities, we note that the summaries remain constrained to the sentence tokens as instructed in the prompts.

\section{Threats to Validity}\label{sec:threats_validity}

\noindent \textbf{Internal Validity} concerns whether the inferences drawn from the experiments are valid. LLMs are generative models; thus, the responses generated by repeating the same prompt may vary. We assess the consistency of F1 scores by repeating each prompt 10 times. Our findings indicate that the variance in F1 scores across different prompts is insignificant. 
We also conducted a manual evaluation study and identified various reasons why the generated summary did not align with the summary templates. This insight can inform modifications to prompts for the summarization task. 
% Also, not all participants of the survey were aware of RoPA thereby making for some instances affecting the quality of the scenarios.  
Additionally, the survey participants and the quality of the scenarios can affect the validity of the generated RoPA segments. 
Finally, further evaluation is required to analyze the usability \& effectiveness of our framework in real-world industry examples.  

\noindent\textbf{External Validity} concerns the extent to which the results generalize beyond the experimental setting. While we exclusively analyzed GPT-3.5 Turbo as an instance of LLMs, further experiments are necessary to compare our findings with other LLMs, such as LLaMA and T5. In addition, GPT-3.5 Turbo undergoes updates in every few months. These updates can also significantly alter the model output over time. Apart from different LLMs and their updates, the use of fine-tuning instead of few-shot learning may yield different results. Further, the generated summaries rely on an input sentence where the action verb is tagged with a specific token. However, further analysis and NLP models are required to identify and label action verbs in the scenario sentences. In this work, we only focus on the effect of the number of examples in few-shot learning and their order. However, further experiments are required to evaluate the effect of prompt structure and instruction on the results.

\noindent\textbf{Reliability} indicates that the researchers' approach is consistent across different researchers and different projects~\cite{gibbs2018}. To ensure the reliability of our research, we measure the inter-coder agreement using Cohen Kappa~\cite{cohen1960coefficient} for the scenario labeling task.

\section{Conclusion \& Future work}\label{sec:conclusion}
%\mitra[inline]{mitra, add stratified sampling}
In this paper, we propose a framework to generate segments of RoPA, which is required for GDPR compliance. Our framework is tailored to support small app-developing companies in their efforts to comply with GDPR and avoid regulatory fines. 
The framework utilizes usage scenarios that contain processing activities. We demonstrate the efficacy of few-shot learning with GPT-3.5 Turbo for generating summaries of processing activities given usage scenarios. Further analysis is needed to evaluate the fine-tuning strategy, utilization of other open-source LLMs, such as LLaMA and T5, and changes to the prompt structure. %However, further work is required to evaluate the usability and effectiveness of extracted summaries in practice. 
Our framework currently lacks an initial phase for identifying and labeling processing activity action verbs within the usage scenarios. As a result, we plan to extend the framework in the future to incorporate models such as Named Entity Recognition (NER) to identify and label these action verbs accurately. Moreover, we intend to conduct empirical studies to evaluate the effectiveness and practical applicability of our framework in extracting RoPA details. %These studies will provide valuable insights into the performance and utility of our approach in real-world scenarios, further refining and enhancing its capabilities.
%add detail about verb classification for future work. This is also the limitation of this work. It needs an initial phase to identify the verbs. 
\bibliographystyle{IEEEtran}
\bibliography{references,ml,policies,domainModel}

\begin{thebibliography}{10}
\providecommand{\url}[1]{#1}
\csname url@samestyle\endcsname
\providecommand{\newblock}{\relax}
\providecommand{\bibinfo}[2]{#2}
\providecommand{\BIBentrySTDinterwordspacing}{\spaceskip=0pt\relax}
\providecommand{\BIBentryALTinterwordstretchfactor}{4}
\providecommand{\BIBentryALTinterwordspacing}{\spaceskip=\fontdimen2\font plus
\BIBentryALTinterwordstretchfactor\fontdimen3\font minus \fontdimen4\font\relax}
\providecommand{\BIBforeignlanguage}[2]{{%
\expandafter\ifx\csname l@#1\endcsname\relax
\typeout{** WARNING: IEEEtran.bst: No hyphenation pattern has been}%
\typeout{** loaded for the language `#1'. Using the pattern for}%
\typeout{** the default language instead.}%
\else
\language=\csname l@#1\endcsname
\fi
#2}}
\providecommand{\BIBdecl}{\relax}
\BIBdecl

\bibitem{wang2018guileak}
X.~Wang, X.~Qin, M.~B. Hosseini, R.~Slavin, T.~D. Breaux, and J.~Niu, ``Guileak: Tracing privacy policy claims on user input data for android applications,'' in \emph{Proceedings of the 40th International Conference on Software Engineering}, 2018, pp. 37--47.

\bibitem{slavin2016toward}
R.~Slavin, X.~Wang, M.~B. Hosseini, J.~Hester, R.~Krishnan, J.~Bhatia, T.~D. Breaux, and J.~Niu, ``Toward a framework for detecting privacy policy violations in android application code,'' in \emph{Proceedings of the 38th International Conference on Software Engineering}, 2016, pp. 25--36.

\bibitem{gdpr}
\BIBentryALTinterwordspacing
{European Parliament and Council of the European Union}, ``General data protection regulation ({GDPR}),'' pp. 1--88, May 2016, originally published on April 27, 2016. [Online]. Available: \url{https://gdpr-info.eu/}
\BIBentrySTDinterwordspacing

\bibitem{huth2019using}
D.~Huth, A.~Tanakol, and F.~Matthes, ``Using enterprise architecture models for creating the record of processing activities (art. 30 gdpr),'' in \emph{2019 IEEE 23rd EDOC}.\hskip 1em plus 0.5em minus 0.4em\relax IEEE, 2019, pp. 98--104.

\bibitem{gdprFines}
``Biggest gdpr fines,'' \url{https://www.tessian.com/blog/biggest-gdpr-fines-2020/}.

\bibitem{balebako2014improving}
R.~Balebako and L.~Cranor, ``Improving app privacy: Nudging app developers to protect user privacy,'' \emph{IEEE Security \& Privacy}, vol.~12, no.~4, pp. 55--58, 2014.

\bibitem{balebako2014privacy}
R.~Balebako, A.~Marsh, J.~Lin, J.~I. Hong, and L.~F. Cranor, ``The privacy and security behaviors of smartphone app developers,'' 2014.

\bibitem{Alomar2022}
S.~E. Noura Alomar~and and J.~L. Fischer, ``Developers say the darnedest things: Privacy compliance processes followed by developers of child-directed apps,'' \emph{Proceedings on Privacy Enhancing Technologies}, vol. 2022, no.~4, 2022.

\bibitem{wagenaar2018working}
G.~Wagenaar, S.~Overbeek, G.~Lucassen, S.~Brinkkemper, and K.~Schneider, ``Working software over comprehensive documentation--rationales of agile teams for artefacts usage,'' \emph{Journal of software engineering research and development}, vol.~6, pp. 1--23, 2018.

\bibitem{hess2017towards}
A.~Hess, P.~Diebold, and N.~Seyff, ``Towards requirements communication and documentation guidelines for agile teams,'' in \emph{2017 ieee 25th international requirements engineering conference workshops (rew)}.\hskip 1em plus 0.5em minus 0.4em\relax IEEE, 2017, pp. 415--418.

\bibitem{Bednar2019}
K.~Bednar, S.~Spiekermann, and M.~Langheinrich, ``Engineering privacy by design: Are engineers ready to live up to the challenge?'' \emph{The Information Society}, vol.~35, no.~3, pp. 122--142, 2019.

\bibitem{Hadar2018}
I.~Hadar, T.~Hasson, O.~Ayalon, E.~Toch, M.~Birnhack, S.~Sherman, and A.~Balissa, ``Privacy by designers: Software developers' privacy mindset,'' \emph{Journal of Empirical Software Engineering}, vol.~23, no.~1, p. 259–289, Feb. 2018.

\bibitem{ekambaranathan2021money}
A.~Ekambaranathan, J.~Zhao, and M.~Van~Kleek, ``“money makes the world go around”: Identifying barriers to better privacy in children’s apps from developers’ perspectives,'' in \emph{Proceedings of the 2021 CHI}, 2021, pp. 1--15.

\bibitem{spiekermann2018inside}
S.~Spiekermann, J.~Korunovska, and M.~Langheinrich, ``Inside the organization: Why privacy and security engineering is a challenge for engineers,'' \emph{Proceedings of the IEEE}, vol. 107, no.~3, pp. 600--615, 2018.

\bibitem{spiekermann2018understanding}
S.~Spiekermann-Hoff, J.~Korunovska, and M.~Langheinrich, ``Understanding engineers' drivers and impediments for ethical system development: The case of privacy and security engineering,'' 2018.

\bibitem{Dalela2021AMS}
A.~Dalela, S.~Giallorenzo, O.~Kulyk, J.~Mauro, and E.~Paja, ``A mixed-method study on security and privacy practices in danish companies,'' \emph{ArXiv}, vol. abs/2104.04030, 2021.

\bibitem{tahaei2021developers}
M.~Tahaei and K.~Vaniea, ``“developers are responsible”: What ad networks tell developers about privacy,'' in \emph{Extended Abstracts of the 2021 CHI}, 2021, pp. 1--11.

\bibitem{Green2016DevelopersAN}
M.~Green and M.~Smith, ``Developers are not the enemy!: The need for usable security apis,'' \emph{IEEE Security \& Privacy}, vol.~14, pp. 40--46, 2016.

\bibitem{Tahaei2021Springer}
M.~Tahaei, A.~Jenkins, K.~Vaniea, and M.~Wolters, ````i don't know too much about it'': On the security mindsets of computer science students,'' in \emph{Socio-Technical Aspects in Security and Trust}, T.~Gro{\ss} and T.~Tryfonas, Eds.\hskip 1em plus 0.5em minus 0.4em\relax Springer International Publishing, 2021.

\bibitem{prybylo2024evaluating}
M.~Prybylo, S.~Haghighi, S.~T. Peddinti, and S.~Ghanavati, ``Evaluating privacy perceptions, experience, and behavior of software development teams,'' 2024.

\bibitem{TahaeiLiVaniea+2022+114+131}
\BIBentryALTinterwordspacing
M.~Tahaei, T.~Li, and K.~Vaniea, ``Understanding privacy-related advice on stack overflow,'' \emph{Proceedings on PETs}, vol. 2022, no.~2, pp. 114--131, 2022. [Online]. Available: \url{https://doi.org/10.2478/popets-2022-0038}
\BIBentrySTDinterwordspacing

\bibitem{RoPAReport}
``Castlebridge register of processing activities (2020),'' \url{https://castlebridge.ie/registers-of-processing-activities-research/}.

\bibitem{martinez2021ontoropa}
M.~M. Mart{\'\i}nez~Gonz{\'a}lez, M.~L. Alvite~D{\'\i}ez, P.~Casanovas, N.~Casellas, D.~Sanz, A.~Aparicio de~la Fuente \emph{et~al.}, ``Ontoropa deliverable 1. state of the art and ambition.'' 2021.

\bibitem{ryan2022support}
P.~Ryan and R.~Brennan, ``Support for enhanced gdpr accountability with the common semantic model for ropa (csm-ropa),'' \emph{SN Computer Science}, vol.~3, no.~3, p. 224, 2022.

\bibitem{anton1994goal}
A.~I. Ant{\'o}n, W.~M. McCracken, and C.~Potts, ``Goal decomposition and scenario analysis in business process reengineering,'' in \emph{Advanced Information Systems Engineering}.\hskip 1em plus 0.5em minus 0.4em\relax Springer, 1994, pp. 94--104.

\bibitem{lubars1993developing}
M.~Lubars, C.~Potts, and C.~Richter, ``Developing initial ooa models,'' in \emph{Proceedings of 1993 15th International Conference on Software Engineering}.\hskip 1em plus 0.5em minus 0.4em\relax IEEE, 1993, pp. 255--264.

\bibitem{voigt2017eu}
P.~Voigt and A.~Von~dem Bussche, ``The eu general data protection regulation (gdpr),'' \emph{A Practical Guide, 1st Ed., Cham: Springer International Publishing}, vol.~10, no. 3152676, pp. 10--5555, 2017.

\bibitem{lankhorst2009enterprise}
M.~Lankhorst \emph{et~al.}, \emph{Enterprise architecture at work}.\hskip 1em plus 0.5em minus 0.4em\relax Springer, 2009, vol. 352.

\bibitem{herwanto2021named}
G.~B. Herwanto, G.~Quirchmayr, and A.~M. Tjoa, ``A named entity recognition based approach for privacy requirements engineering,'' in \emph{2021 IEEE 29th RE Workshops (REW)}.\hskip 1em plus 0.5em minus 0.4em\relax IEEE, 2021, pp. 406--411.

\bibitem{Sleimi2018AutomatedEO}
A.~Sleimi, N.~Sannier, M.~Sabetzadeh, L.~Briand, and J.~Dann, ``Automated extraction of semantic legal metadata using natural language processing,'' \emph{2018 IEEE 26th RE}, pp. 124--135, 2018.

\bibitem{sleimi2020}
A.~Sleimi, M.~Ceci, M.~Sabetzadeh, L.~C. Briand, and J.~Dann, ``Automated recommendation of templates for legal requirements,'' in \emph{2020 IEEE 28th RE}, 2020, pp. 158--168.

\bibitem{CEJAS2021}
O.~Amaral~CEJAS, S.~Abualhaija, D.~Torre, M.~Sabetzadeh, and L.~Briand, ``Ai-enabled automation for completeness checking of privacy policies,'' \emph{IEEE Transactions on Software Engineering}, pp. 1--1, 2021.

\bibitem{Amaral2021}
O.~Amaral, S.~Abualhaija, M.~Sabetzadeh, and L.~Briand, ``A model-based conceptualization of requirements for compliance checking of data processing against gdpr,'' in \emph{2021 IEEE 29th RE Workshops (REW)}, 2021, pp. 16--20.

\bibitem{sabetzadeh21}
S.~Ezzini, S.~Abualhaija, C.~Arora, M.~Sabetzadeh, and L.~Briand, ``Maana: An automated tool for domain-specific handling of ambiguity,'' in \emph{2021 IEEE/ACM 43rd ICSE-Companion}, 2021, pp. 188--189.

\bibitem{herwanto2022privacystory}
G.~B. Herwanto, G.~Quirchmayr, and A.~M. Tjoa, ``Privacystory: Tool support for extracting privacy requirements from user stories,'' in \emph{2022 IEEE 30th RE}.\hskip 1em plus 0.5em minus 0.4em\relax IEEE, 2022, pp. 264--265.

\bibitem{Gha13}
S.~Ghanavati, ``Legal-{{URN}} framework for legal compliance of business processes,'' Ph.D. dissertation, University of Ottawa, Ottawa, Canada, 2013.

\bibitem{GRD14}
S.~Ghanavati, A.~Rifaut, E.~Dubois, and D.~Amyot, ``Goal-oriented compliance with multiple regulations,'' in \emph{2014 {{IEEE}} 22nd {{{RE}}}}, 2014, pp. 73--82.

\bibitem{anton1998representational}
A.~I. Ant{\'o}n and C.~Potts, ``A representational framework for scenarios of system use,'' \emph{Requirements Engineering}, vol.~3, pp. 219--241, 1998.

\bibitem{sutcliffe1998scenario}
A.~Sutcliffe, ``Scenario-based requirements analysis,'' \emph{Requirements engineering}, vol.~3, pp. 48--65, 1998.

\bibitem{sutcliffe2003scenario}
------, ``Scenario-based requirements engineering,'' in \emph{Proceedings. 11th IEEE RE, 2003.}\hskip 1em plus 0.5em minus 0.4em\relax IEEE, 2003, pp. 320--329.

\bibitem{weidenhaupt1998scenarios}
K.~Weidenhaupt, K.~Pohl, M.~Jarke, and P.~Haumer, ``Scenarios in system development: current practice,'' \emph{IEEE software}, vol.~15, no.~2, pp. 34--45, 1998.

\bibitem{lubars1993review}
M.~Lubars, C.~Potts, and C.~Richter, ``A review of the state of the practice in requirements modeling,'' in \emph{IEEE RE}.\hskip 1em plus 0.5em minus 0.4em\relax IEEE, 1993, pp. 2--14.

\bibitem{potts1994inquiry}
C.~Potts, K.~Takahashi, and A.~I. Anton, ``Inquiry-based requirements analysis,'' \emph{IEEE software}, vol.~11, no.~2, pp. 21--32, 1994.

\bibitem{huang2023mobile}
T.~Huang, V.~Kaulagi, M.~B. Hosseini, and T.~Breaux, ``Mobile application privacy risk assessments from user-authored scenarios,'' in \emph{2023 IEEE 31st RE}.\hskip 1em plus 0.5em minus 0.4em\relax IEEE, 2023, pp. 17--28.

\bibitem{filatova2004event}
E.~Filatova and V.~Hatzivassiloglou, ``Event-based extractive summarization,'' 2004.

\bibitem{hsu2018unified}
W.-T. Hsu, C.-K. Lin, M.-Y. Lee, K.~Min, J.~Tang, and M.~Sun, ``A unified model for extractive and abstractive summarization using inconsistency loss,'' in \emph{ACL}, 2018, pp. 132--141.

\bibitem{lin2019abstractive}
H.~Lin and V.~Ng, ``Abstractive summarization: A survey of the state of the art,'' in \emph{Proceedings of the AAAI conference on artificial intelligence}, vol.~33, no.~01, 2019, pp. 9815--9822.

\bibitem{jadhav2018extractive}
A.~Jadhav and V.~Rajan, ``Extractive summarization with swap-net: Sentences and words from alternating pointer networks,'' in \emph{ACL 2018-56th}, vol.~1.\hskip 1em plus 0.5em minus 0.4em\relax Association for Computational Linguistics (ACL), 2018, pp. 142--151.

\bibitem{nallapati2017summarunner}
R.~Nallapati, F.~Zhai, and B.~Zhou, ``Summarunner: A recurrent neural network based sequence model for extractive summarization of documents,'' in \emph{Proceedings of the AAAI conference on artificial intelligence}, vol.~31, no.~1, 2017.

\bibitem{yadav2024graph}
A.~K. Yadav, Ranvijay, R.~S. Yadav, and A.~K. Maurya, ``Graph-based extractive text summarization based on single document,'' \emph{Multimedia Tools and Applications}, vol.~83, no.~7, pp. 18\,987--19\,013, 2024.

\bibitem{mishra2023llm}
N.~Mishra, G.~Sahu, I.~Calixto, A.~Abu-Hanna, and I.~Laradji, ``Llm aided semi-supervision for efficient extractive dialog summarization,'' in \emph{Findings of the Association for Computational Linguistics: EMNLP 2023}, 2023, pp. 10\,002--10\,009.

\bibitem{fantechi2023rule}
A.~Fantechi, S.~Gnesi, and L.~Semini, ``Rule-based nlp vs chatgpt in ambiguity detection, a preliminary study,'' 2023.

\bibitem{lami2004automatic}
G.~Lami, S.~Gnesi, F.~Fabbrini, M.~Fusani, and G.~Trentanni, ``An automatic tool for the analysis of natural language requirements,'' \emph{Informe t{\'e}cnico, CNR Information Science and Technology Institute, Pisa, Italia, Setiembre}, 2004.

\bibitem{ruan2023requirements}
K.~Ruan, X.~Chen, and Z.~Jin, ``Requirements modeling aided by chatgpt: An experience in embedded systems,'' in \emph{2023 IEEE 31st RE (REW)}.\hskip 1em plus 0.5em minus 0.4em\relax IEEE, 2023, pp. 170--177.

\bibitem{gorer2023generating}
B.~G{\"o}rer and F.~B. Aydemir, ``Generating requirements elicitation interview scripts with large language models,'' in \emph{2023 IEEE 31st RE (REW)}.\hskip 1em plus 0.5em minus 0.4em\relax IEEE, 2023, pp. 44--51.

\bibitem{jackson1995software}
M.~Jackson, \emph{Software Requirements \& Specifications: a lexicon of practice, principles and prejudices}.\hskip 1em plus 0.5em minus 0.4em\relax ACM Press/Addison-Wesley Publishing Co., 1995.

\bibitem{cohen1960coefficient}
J.~Cohen, ``A coefficient of agreement for nominal scales,'' \emph{Educational and psychological measurement}, vol.~20, no.~1, pp. 37--46, 1960.

\bibitem{landis1977measurement}
J.~R. Landis and G.~G. Koch, ``The measurement of observer agreement for categorical data,'' \emph{biometrics}, pp. 159--174, 1977.

\bibitem{chen2023unleashing}
B.~Chen, Z.~Zhang, N.~Langren{\'e}, and S.~Zhu, ``Unleashing the potential of prompt engineering in large language models: a comprehensive review,'' \emph{arXiv preprint arXiv:2310.14735}, 2023.

\bibitem{yu2023exploring}
F.~Yu, L.~Quartey, and F.~Schilder, ``Exploring the effectiveness of prompt engineering for legal reasoning tasks,'' in \emph{Findings of the Association for Computational Linguistics: ACL 2023}, 2023, pp. 13\,582--13\,596.

\bibitem{oleaevaluating}
C.~Olea, H.~Tucker, J.~Phelan, C.~Pattison, S.~Zhang, M.~Lieb, D.~Schmidt, and J.~White, ``Evaluating persona prompting for question answering tasks.''

\bibitem{white2023prompt}
J.~White, Q.~Fu, S.~Hays, M.~Sandborn, C.~Olea, H.~Gilbert, A.~Elnashar, J.~Spencer-Smith, and D.~C. Schmidt, ``A prompt pattern catalog to enhance prompt engineering with chatgpt,'' \emph{arXiv preprint arXiv:2302.11382}, 2023.

\bibitem{akter2022revisiting}
M.~Akter, N.~Bansal, and S.~K. Karmaker, ``Revisiting automatic evaluation of extractive summarization task: Can we do better than rouge?'' in \emph{ACL 2022}, 2022, pp. 1547--1560.

\bibitem{nanba2006automatic}
H.~Nanba and M.~Okumura, ``An automatic method for summary evaluation using multiple evaluation results by a manual method,'' in \emph{COLING/ACL}, 2006, pp. 603--610.

\bibitem{banerjee2005meteor}
S.~Banerjee and A.~Lavie, ``Meteor: An automatic metric for mt evaluation with improved correlation with human judgments,'' in \emph{Proceedings of the acl workshop on intrinsic and extrinsic evaluation measures for machine translation and/or summarization}, 2005, pp. 65--72.

\bibitem{soleimani2023nonfacts}
A.~Soleimani, C.~Monz, and M.~Worring, ``Nonfacts: Nonfactual summary generation for factuality evaluation in document summarization,'' in \emph{ACL 2023}, 2023, pp. 6405--6419.

\bibitem{gibbs2018}
G.~R. Gibbs, \emph{Analyzing qualitative data}.\hskip 1em plus 0.5em minus 0.4em\relax Sage, 2018, vol.~6.

\end{thebibliography}

\end{document}